\documentclass[12pt,preprint]{aastex}

\usepackage{natbib}

\shorttitle{Coronal Magnetography of a Simulated Solar Active Region}
\shortauthors{Wang et al.}

\begin{document}


\title{Coronal Magnetography of a Simulated Solar Active Region from Microwave Imaging Spectropolarimetry}


\author{Zhitao Wang\altaffilmark{1}, Dale E. Gary\altaffilmark{1}, Gregory D. Fleishman\altaffilmark{1}, and Stephen M. White\altaffilmark{2}}

\affil{1. Center for Solar-Terrestrial Research, New Jersey Institute of Technology, University Heights, Newark, NJ 07102}

\affil{2. Air Force Research Laboratory, Albuquerque, NM, USA}
\email{dgary@njit.edu}




\begin{abstract}
We have simulated the Expanded Owens Valley Solar Array (EOVSA) radio images generated at multiple frequencies from a model solar active region, embedded in a realistic solar disk model, and explored the resulting datacube for different spectral analysis schemes to evaluate the potential for realizing one of EOVSA's most important scientific goals\textemdash{coronal magnetography}. In this paper, we focus on modeling the gyroresonance and free-free emission from an on-disk solar active region model with realistic complexities in electron density, temperature and magnetic field distribution. We compare the magnetic field parameters extrapolated from the image datacube along each line of sight after folding through the EOVSA instrumental profile with the original (unfolded) parameters used in the model. We find that even the most easily automated, image-based analysis approach (Level 0) provides reasonable quantitative results, although they are affected by systematic effects due to finite sampling in the Fourier (uv) plane. Finally, we note the potential for errors due to misidentified harmonics of the gyrofrequency, and discuss the prospects for applying a more sophisticated spectrally-based analysis scheme (Level 1) to resolve the issue in cases where improved uv coverage and spatial resolution are available.

\end{abstract}





\section{Introduction}
The magnetic field in the solar atmosphere plays a critical role in determining the plasma structure, storage of free magnetic energy, and its release in various forms of solar activity, such as flares and coronal mass ejections (CMEs). In order to interpret such solar phenomena, the calculation of free magnetic energy and its storage-release process requires quantitative knowledge of the three-dimensional (3D) structure of the corona in or near the span of eruption.

One way to derive the coronal field strength is through nonlinear force-free field (NLFFF) extrapolations from the measured photospheric boundary data. To further verify the model, morphological tests between the simulated field lines and 2D coronal brightness maps observed in EUV and soft X-rays are performed. However, these extrapolations are found to be imperfect when applied to real solar data: different algorithms demonstrate various coronal field configurations and show profound differences in the magnetic energy \citep{2008ApJ...675.1637S, 2009ApJ...696.1780D}. In addition, since NLFFF algorithms are very sensitive to the force-free boundary data \citep{2006SoPh..235..161S, 2008SoPh..247..269M}, several major factors which may affect the boundary become critical to the success of NLFFF modeling in real applications: (1) measurement uncertainties, such as noise levels and the 180-degree ambiguity, (2) measurement limitations, specifically deriving the vector magnetic field over a field of view large enough to accommodate most of the field-line connectivity within the active region and its surroundings, and insufficient spatial resolution to recover the distribution of current densities on the boundary, and (3) poor knowledge of an appropriate procedure for transforming the non-force-free photospheric magnetic fields into an approximately force-free boundary. Even if some of these issues can be addressed, NLFFF extrapolation relies heavily on the assumption that the coronal field is force-free, which is not always true in the case of fast eruptive activities. Moreover, morphological tests are constrained by the 2D observational clues to the coronal structure, where the underlying field lines may not necessarily follow the observed loop structure in brightness. Therefore, it is essential to develop direct measurement techniques in conjunction with extrapolations to better determine the coronal magnetic field.

Unfortunately, it is very difficult to directly measure the coronal magnetic field through conventional Zeeman sensitive spectral lines. On the disk, the bright photospheric and chromospheric emission generally overwhelms the coronal contribution, so Zeeman studies have been carried out above the limb, where line-of-sight integration may complicate the interpretation. Since Zeeman sensitivity increases at longer wavelengths, many studies in recent years have focussed on the promising application of infrared (IR) coronal emission lines (CELs) \citep{1998ApJ...500.1009J, 2000ApJ...541L..83L, 2004ApJ...613L.177L}. Although direct comparison between the measurement of CELs and coronal magnetic field models has been attempted \citep{2008ApJ...680.1496L}, the reliability of both the measurement and the model remain to be validated.

It has been recognized that the use of microwave imaging spectroscopy can provide prominent diagnostics of the coronal magnetic field through the spectral analysis of two radiative mechanisms: gyroresonance emission and free-free emission \citep{1994ApJ...420..903G, 1999SoPh..185..157R,2000A&AS..144..169G}. During non-flaring times in the lower corona above active regions, where magnetic fields are strong (typically in the range 100 to 2000 G), gyroresonance emission dominates in the 1-20 GHz range of microwave frequencies. The polarization and the spectra of gyroresonance emission provide the diagnostic information needed to derive coronal magnetic fields, which has a long history \citep{1968AnAp...31..105L, 1968aSvA....12..245Z, 1968bSvA....12..464Z, 1979SvA....23..316G, 1980A&A....82...30A, 1982SoPh...79...41A, 1984A&A...139..271A, 1986ApJ...301..460A, 2006ARep...50..679P, 2009AstBu..64..372B, 2012ARep...56..790K}. See also reviews by \citet{1997SoPh..174...31W, 2004ASSL..314.....G, 2007SSRv..133...73L}. However, so far there remains no comprehensive test of the reliability of this diagnostic method for use on a daily basis in practical observations. With the recent advances in radio interferometric instruments that will soon be capable of combining adequate spatial and spectral resolution for coronal magnetic field diagnostics, we provide a systematic evaluation of this approach, using the Expanded Owens Valley Solar Array (EOVSA) instrument profile as a specific example for quantitative comparison of the folded parameters with the original model.

\section{Modeling Background}

\subsection{The Origin of the Mok Model}\label{mokmodel}
To present a realistic challenge in both the imaging and spectral domains, we exploit a physically plausible model from \citet{2005ApJ...621.1098M}, which includes a spatial data cube of electron density $n_e$, electron temperature $T_e$, and vector magnetic field $\textbf{B}$. Calculation of these parameters follows two major steps: (1) the potential field model is first applied to extrapolate the magnetic field specification from the vector magnetogram data of one selected active region; (2) after the field line configuration is determined, the thermal structure of the corona is computed self-consistently using a steady state heating model in which the volumetric heating rate is simplified as directly proportional to the static magnetic field strength. The original data used by \citet{2005ApJ...621.1098M} was based on active region 7986, a spatially-dispersed, magnetically weak active region observed during solar minimum in August 1996. To simulate a more typical active region in the corona, we reduced the spatial pixel dimension from $6.2\times6.8\arcsec$ to $2.4\times2.4\arcsec$, and scaled the total field of view (FOV) from Mok's model to be $172\arcsec\times127\arcsec$, while keeping the original scale height. We also scale the magnetic field, whose original longitudinal component ranges from $-$221~G to 179~G, by a factor of 10, and reduce the electron density by 33\%, to match typical ranges in the corona above strong active regions. Although these changes invalidate the physical correctness of the active region model, for our purposes this is unimportant.  We require only that the model contains sufficiently realistic complexity to be a fair test of the ability of extracting the coronal plasma parameters of the model from the simulated multifrequency radio images. Although the spatial resolution of the model is too low to reveal distinct loop-like features, it contains a physical resemblance to the actual solar corona, and for our purpose provides sufficient complexity for a practical evaluation. Figure~\ref{fig1} shows the 2D distribution of the longitudinal magnetic field, the electron density, and the temperature, including the cross-section view at the base of the corona, and at the horizontal center of the model. We follow the nomenclature in \citet{2013SoPh..288..549G} and refer to the direct images calculated from the model as "unfolded" images; and those obtained after sampling with the spatial response of EOVSA as "folded" images, ie. folded through the instrument response function. We continue to discuss the use of the model to generate the images at multiple microwave frequencies in section 3.

\subsection{Emission Mechanisms and Radiation Transfer}\label{emissionmech}
We now briefly review two dominant mechanisms, gyroresonance and free-free emission, which have been implemented in our source code to simulate multifrequency images (a data cube) from the original model. Gyroresonance emission depends strongly on the plasma temperature (and, more broadly, the distribution function type), magnetic field strength and direction, and the low harmonics of gyrofrequency $\nu_B \approx 2.8\times10^6 B\,{\rm Hz}$ are known to match the range of microwave frequency (1-20 GHz) in the corona where the magnetic field strengths can reach up to 2500 Gauss. Although the theory developed by \citet{2014ApJ...781...77F} allows non-Maxwellian distributions, in this paper we consider the Maxwellian distribution  only. For this case, the absorption coefficient due to a given gyro harmonic $s$ has the form (e.g., \citet{1970resp.book.....Z}):

\begin{eqnarray}
\label{kappa_gr}
\kappa_{gr,\sigma} =\frac{\sqrt{2\pi}e^2n_{e}c}{\nu k_{B}T}{\left(\frac{k_{B}T}{m_ec^2}\right)}^{s-1/2}
\frac{s^{2s} n_{\sigma}^{2s-4}\sin^{2s-2}\theta}{2^s s!\left(1+T_{\sigma}^2\right)|\cos\theta|}\times[T_{\sigma}\cos\theta+L_{\sigma}\sin\theta+1]^2 \nonumber
\end{eqnarray}

\begin{eqnarray}
 \times \exp\left\{-\frac{m_ec^2}{2k_{B}T}\frac{\left(\nu-s\nu_{B}\right)^2}{\nu^{2}n_{\sigma}^{2}\cos^2\theta}\right\},
\end{eqnarray}
from which the emissivity can be straightforwardly found using the Kirchhoff law
\begin{equation}
\label{Eq_Kirg_gr}
 j^{gr}_{\sigma,\nu}\approx
 \frac{n_\sigma^2\nu^2}{ c^2 }k_BT \kappa_{gr,\sigma},
\end{equation}
where $B$ is the total magnetic field strength, $e$ is the electron charge, $m_e$ is the electron mass, $n_{\sigma}$ is the refractive index of the wave-mode $\sigma$, $\theta$ is the angle of the field to the line of sight, $T_{\sigma}$ and $L_{\sigma}$ are the components of the wave polarization vector (all variable definitions follow the conventions  adopted in \citet{2010ApJ...721.1127F, 2014ApJ...781...77F}).
Note that the presence of the exponential factor in Eq~(\ref{kappa_gr}) makes the gyro opacity exponentially small everywhere beyond a very narrow range of frequencies very close to the gyro harmonics $s\nu_{B}$. For emission produced at a given frequency in the solar atmosphere, where the magnetic field varies with height, this narrowness translates to a narrow range of heights, called a \textit{gyrolayer}, which makes a dominant contribution to the opacity at this frequency. For typical conditions of the corona, where the magnetic field obeys the resonant condition
\begin{equation}
\label{gr}
\nu = s\nu_B = s {{eB}\over{2\pi m_e c}} \approx 2.8\times 10^{6}sB\,\ \ {\rm Hz},
\end{equation}
the gyrolayers are only about 100 km thick. For the solar corona, $s$ is a small integer, typically 1, 2 or 3, although in some cases s = 4 may have appreciable opacity \citep{1997SoPh..174...31W}.

Whenever the gyroresonance emission is small, the free-free emission makes a dominant contribution \citep{2004ASSL..314.....G}. In a magnetized plasma the free-free emission of a given eigen-mode has the form \citep{1968aSvA....12..245Z}\footnote{Expression of the factor $F_\sigma$ responsible (along with $n_{\sigma}$ in the denominator) for polarization of the free-free emission is given in textbook by \citet{2013ASSL..388.....F} but with a typo resulted in the wrong general sign of the expression.}:
\begin{equation}
\label{kappa_ff}
\kappa_{ff,\sigma}=\frac{8e^6\ln{\Lambda_{C}}}{3\sqrt{2\pi}n_{\sigma}c\nu^2(m_ek_BT)^{3/2}}\times n_{e}(n_{II} + 4n_{HeIII})F_\sigma
\end{equation}
where $n_{II}$ is the total number density of all singly ionized atoms (mainly--protons)\footnote{If no independent information about ionization states (e.g., non-LTE) is supplied by the given model, the code computes the ionization states of hydrogen and helium using the Saha equation based on the electron density and temperature input. Note that for the coronal temperatures and coronal abundances, $n_{e}(n_{II} + 4n_{HeIII})\approx 1.14 n_e^2$.}, $n_{II}\approx n_{e} - 2n_{HeIII}$, and $n_{HeIII}$ is the number density of the fully ionized Helium
\begin{equation}
\label{F_sigma}
 F_\sigma = 2~\frac{
 \sigma\sqrt{\mathcal{D}}\left[u\sin^2\theta+2(1-v)^2\right] - u^2\sin^4\theta }
 {\sigma\sqrt{\mathcal{D}}\left[2(1-v)-u\sin^2\theta+\sigma\sqrt{\mathcal{D}}\right]^2}
 ,
\end{equation}
\begin{equation}
\mathcal{D}=u^2\sin^4\theta+4u(1-v)^2\cos^2\theta,
\end{equation}
\begin{equation}
u=\left(\frac{\nu_B}{\nu}\right)^2,\qquad
v=\left(\frac{\nu_{\mathrm{pe}}}{\nu}\right)^2,
\end{equation}
$\nu_{\mathrm{pe}}=e\sqrt{n_{e}/(\pi m_{\mathrm{e}})}$ is the electron plasma
frequency. For X-mode, $\sigma=-1$; for O-mode, $\sigma=+1$,
$\Lambda_{C}$ is the Coulomb logarithm.

Unlike the case of gyroresonance opacity, the free-free opacity is generated over a much larger LOS, although when seen optically thick against the solar disk it is heavily weighted toward the lowest, densest part of the ray path; the lower the frequency, the higher the effective formation height \citep[e.g.,][]{2015arXiv150102898L} of the free-free emission ranging from chromosphere at high frequencies ($\gtrsim 10$~GHz) to corona at lower frequencies.

In principle, the resulting gyroresonance and free-free optical depth and emerging radiation can be calculated for each volume element (voxel) from any model that provides electron number density, electron temperature, and vector magnetic field in each voxel. However, the size of the voxel height is typically much larger than the depth of a gyro layer. For that reason, the computation engine \citep[see the complete description in][]{2014ApJ...781...77F} linearly interpolates the magnetic field vector between the neighboring voxels to precisely locate the positions of all gyro layers of interest (the default maximum value $s_{\max}=7$ is adopted but can easily be adjusted by the user) to calculate the absorption coefficient (\ref{kappa_ff}), and then solves the equation of radiation transfer, taking account of frequency-dependent mode coupling as described in \citet{2014arXiv1409.0896N}. The equation of radiation transfer solves for the intensities of the X and O modes, which have elliptical (i.e., not necessarily circular) polarization when they have just left the data cube. We take into account, however, that the waves continue to propagate through the corona with declining electron density and magnetic field and so the elliptical polarization necessarily evolves towards a truly circular one due to the effect of 'limiting polarization' \citep{1970resp.book.....Z, 2013ASSL..388.....F}; thus, we identify the intensities of the X and O modes with the circularly polarized waves observed at the Earth.

\section{Modeling Scheme}
\subsection{Generation of Unfolded Images}\label{bozomath}
In order to simulate the multifrequency unfolded images, we calculate the radio emission for each LOS pixel, at 64 logarithmically-spaced frequencies in the range 1-18 GHz. By considering each voxel as a homogeneous source, the model parameters at the center of a given voxel can be used to calculate the resulting, frequency-dependent total optical depth $\tau_{t,\sigma}=\tau_{gr,\sigma}+\tau_{ff,\sigma}$ in two polarizations at that voxel. Subsequently, the total brightness temperature of a given pixel at each frequency can be obtained through appropriate integration of radiative transfer in the corresponding voxels along the LOS. These calculations are routinely done within the GX\textunderscore simulator tool \citep{2014arXiv1409.0896N} to produce a full set of simulated 2D brightness temperature maps at the chosen frequencies. The first column from left of Figure~\ref{fig2} shows a sample of 6 of these RCP (unfolded) images out of the 64-frequency data cube. The similar LCP (unfolded) images are also shown in the third column of Figure~\ref{fig2}.

\subsection{Generation of the Folded Images}
In an actual observation of gyroresonance emission, from which the coronal magnetic field is to be inferred, the target active region brightness is competing with that of other active regions and the radio-bright solar disk, so that the uv visibilities contain contributions from this complex spatial structure. This can contribute to confusion in the reconstructed image due to finite sampling in the $uv$ plane. This problem is especially true in EOVSA's case, since it consists of a limited number of small antennas (a total of 13 2.1-m antennas). At 18~GHz, the primary beam size of one EOVSA antenna is about 33~arcmin, increasing to 18 times larger at 1 GHz following the diffraction limit. Therefore, to provide a fair test of the instrument performance, we embed the unfolded images, without loss of generality, into the central region of a full-disk model. Figure~\ref{fig3} is one of the 128 simulated full disk images (64 frequencies and 2 polarizations) at one selected frequency and polarization.

The frequency--dependent full-disk model is generated as follows. As a base map we use SOHO/EIT EUV images at 171  and 195 \AA. The ratio of these two images is converted to emission measure and temperature (in the range 0.6--1.7 MK) using standard EIT software and assuming a coronal abundance for Fe. The date of the EUV images is 1999 Apr 11 (at 19 UT), chosen because it was available from an earlier (unpublished) study. From the emission--measure and temperature maps, we can generate a model radio image (in units of brightness temperature for convenience) from bremsstrahlung that correctly accounts for optical depth variation with frequency (i.e., at low frequencies active regions are optically thick at the temperature of the model).

The EIT model only reproduces bremsstrahlung from coronal plasma. The radio images also see a background from the cooler lower atmosphere that we add as a frequency--dependent disk of a brightness temperature that results from fits to the solar radio spectrum at solar minimum, assuming an effective disk radius determined from fits to the size of full--disk images at frequencies where such images are available \citep[e.g.][---at 4.5 GHz the radio limb is 30\arcsec\ above the photosphere]{1996ASPC...93..387G}.

Gyroresonance sources typically lie above strong magnetic fields in the photosphere. In order to simulate them we use an MDI line--of--sight magnetogram, and apply a frequency--dependent threshold in field strength as well as a dilation transformation that gives gyroresonance sources of about the same size as are seen in the actual radio images on 1999 April 11. The gyroresonance sources are assumed to be optically thick at 2.5 MK. As an ansatz to generate plausible polarization models, gyroresonance sources with positive polarity are added to the right circular polarization model (replacing the other sources, since the gyroresonance sources are optically thick), while those with negative polarity are added to the left circular polarization model, i.e., all gyroresonance sources are assumed to be 100\% circularly polarized, which is not true of actual sources. Lastly, for each disk brightness model with suitably realistic spatial complexity, the unfolded images from the active region model at the corresponding frequency are added at a spatial location near disk center, as shown for 4.15 GHz in Figure~\ref{fig3}.

To produce the instrument-folded image datacube, we use EOVSA as a test case, i.e., a 13-antenna array (78 baselines) with imaging from 1 to 18 GHz. Figure~\ref{fig4} shows the $uv$ coverage of EOVSA at 18 GHz, used for a simulation covering 12 hours of observation on a date when the Sun is at +$15^\circ$ declination. Then, an automatic procedure using the \emph{Miriad} package \citep{1995ASPC...77..433S} is followed: (1) the brightness image of the disk with embedded active region model is Fourier-transformed into the uv plane to generate the resulting model visibility function $V(u,v)$; (2) the true visibility function is sampled with the 78 baselines of instrumental profile (the samples at each frequency and time $V(u,v,t,\nu)$ are termed visibilities), shown in Figure~\ref{fig4}. Since the effect of bandwidth smearing can potentially reduce the quality of reconstructed images, we include a bandwidth of 50 MHz in the uv sampling, which slightly exceeds the maximum limit of EOVSA's bandwidth (variable up to 40 MHz); (3) thermal noise from the system, $4500~K$, is added to the visibilities to simulate the effect of the bright solar emission on system temperature, and (4) the image is reconstructed using \emph{Miriad}'s standard Maximum Entropy Method (MEM) routine \emph{maxen}. By repeating (1)-(4) for each frequency and polarization, a full set of RCP and LCP images is generated at all sampled frequencies.

The second and fourth columns of Figure~\ref{fig2}, shown at the same frequencies as in the unfolded images in the first and third columns correspond to the relevant portions of the simulated (folded) images in RCP and LCP, respectively, after being sampled with the uv-coverage of the EOVSA instrument (Figure~\ref{fig4}) . In general, the features of the active region from the folded maps follow the unfolded ones very well, especially in higher frequency range, say above 4 GHz. However, the folded images tend to be lacking in details at lower frequencies: around 1-2 GHz, the original sharp drops in brightness temperature within the opacity holes directly over the sunspot areas are fairly smooth in the folded maps in both polarizations. This is due to the poorer spatial resolution at lower frequencies, as set by the maximal projected baseline $B_{\lambda,proj}$ of the antenna arrays. We show in the following that even with limited visibility coverage, EOVSA can provide reliable spectra for the coronal magnetic field measurement in such a complex active region structure.

\section{Results and Analysis}
\subsection{Comparison of the Brightness Temperature Map and Spectra}{\label{comparison}}

We start with the elementary radiative transfer problem of radio emission and extend our discussion over emission features from our active region model in the unfolded $T_{b}$ maps. In microwaves, the intensity of the source can be approximated by the Rayleigh-Jeans law in terms of $I_\nu\approx2k\nu^2T_b/c^2$, where $T_b$ is the brightness temperature, which is equivalent to the temperature of a blackbody having the same brightness. The radiative transfer equation in terms of brightness temperature is given by:

\begin{equation}
\label{tra}
T_b(\tau)=\int_0^\tau T_e(\tau')e^{-\tau'}d\tau+T_b(0)e^{-\tau}
\end{equation}
where $\tau$ is the maximum opacity along a given LOS. The second term represents the contribution of a background such as the optically-thick chromosphere. For an optical-thick thermal source under thermal equilibrium ($\tau \gg 1$), the observed brightness temperature is approximately equivalent to the local coronal electron temperature $T_b\approx T_e$. For gyroresonance emission alone, which is the dominant mechanism in the strong magnetic field (above 100 Gauss), emission is confined to the harmonic layers satisfying equation (1). Therefore the measured brightness temperature along a given LOS is contributed by the outermost gyroresonance layer that is optically thick at a given observing frequency.

It is instructive to view the unfolded data cube vertically to examine the microwave spectral properties at particular LOS positions, from which the physical parameters can in principle be extracted through spectral analysis. Figure~\ref{fig5}(a)-(b) show the unfolded and folded images in RCP at 4.3 GHz, which corresponds to the magnetic field strength $|B|\approx512$~G of the third harmonic layer. The numbered points indicate the locations of six sampled spectra in Figure~\ref{fig5}(c)-(h). Red and blue represent the right- and left-circular polarized (RCP and LCP) spectra. The unfolded spectra (solid lines) in the two polarizations at a given LOS differ due to differences in the highest optically-thick harmonic, which can be read directly from the unfolded spectra.

To aid in further discussion, we show in Figure~\ref{fig6} the parameters of the model as a function of height along the 6 lines of sight marked in Figure~\ref{fig5}. The inferred brightness temperature of point 1 (red curve), sampled near the loop-top region, is around $T_R\approx2.3\times10^6$ K at 4.3 GHz (Figure~\ref{fig5}c). The height-dependent magnetic field of the model (Figure~\ref{fig6}b) shows that the corresponding 3rd-harmonic field strength $|B|=512$~G (thin dotted lines in Figure~\ref{fig6}b) meets the red, point 1 curve at a coronal height of $H\approx38$~Mm, where the coronal electron temperature is $T_e\approx2.7$~MK (Figure~\ref{fig6}a). At point 3, both RCP and LCP spectra (Figure~\ref{fig5}e) show steep drops in brightness temperature at 6~GHz. The LCP spectrum then maintains a moderate brightness above 6~GHz, with a second drop in brightness around 9 GHz. This corresponds to the next higher ($s=3$) harmonic layer, which is partially optically thin, dropping out of the corona. The 2:3 ratio of the frequencies of the two sharp drops in the spectrum indicates that the turn-over frequency of 6~GHz corresponds to the second harmonic layer. Accordingly, at 6~GHz the field strength derived from $s=2$ is $|B|\approx1071$~G, which is the limiting field strength along that LOS (Figure~\ref{fig6}b), and occurs at a model height of $H\approx1$~Mm and temperature $T_e\approx10^4$~K.  This accounts for the sudden drop in brightness at that frequency. Away from the active region (points 4-6) where the 512~G isogauss layer does not lie in the corona (the curves for points 4-6 all fall below the thin dotted line in Figure~\ref{fig6}b), free-free emission is the dominant mechanism. For example, at point 4 above 3~GHz (Figure~\ref{fig5}f), the gradually-decaying spectrum suggests that bremsstrahlung is the primary emission mechanism.

It is also important to interpret the nature of the sudden depression in the brightness temperature over the sunspot regions in both polarizations (e.g $T_R$ at point 2 and 3, Figure~\ref{fig5}a) in terms of local spectral behavior (Figs.~\ref{fig5}d,e). The reason for the depression is that when the angle $\theta$ between the vector magnetic field and the LOS approaches $0^\circ$ or $180^\circ$, the corresponding gyroresonance layer becomes optically thin. This can occur in the $O$-mode emission at larger angle $\theta=\theta_c$ than for $X$-mode emission \citep{1997SoPh..174...31W,2004ASSL..314.....G}. At point 2, the angle $\theta$ along the LOS (magenta line in Figure~\ref{fig6}c) is less than $20^\circ$, which falls below the critical angle $\theta_c$ in the $O$ mode ($T_R$). Thus, there is no significant brightness contribution to $T_R$ from the optically thin harmonic layers along this LOS, leaving only free-free emission in the $T_R$ spectra (Figure~\ref{fig5}d). In comparison, point 3 (Figure~\ref{fig5}e) also shows similar spectral behavior in the 1-2.5 GHz range, except for a high brightness temperature bump that peaks at 5.2 GHz.  We showed earlier that this corresponds to second-harmonic emission, which yields a field strength $B=928$~G (thick dotted line in Figure~\ref{fig6}b). The corresponding angle in Figure~\ref{fig6}c shows $\theta=33.2^\circ$, which is at least $10^\circ$ larger than at point 2. In this range, $\theta$ approaches $\theta_c$, increasing the opacity and hence the brightness, as frequency increases toward 6 GHz. The variation of brightness temperature with viewing angle is rich in diagnostic power, but does complicate the interpretation of spectra (see section 4.3).

Finally, the dashed lines in Figure~\ref{fig5}c-f show the effect on the spectrum of folding the model images through the instrument profile. Some folded spectra match the unfolded spectra (solid lines), but due to the limited $uv$ coverage and the resulting finite spatial resolution, the folded spectra tend to be both more extended and lacking in spectral details. Excellent agreement can be found in the center of the active region, where the spatial brightness temperature gradient is relatively smooth. For example, the folded spectra from point 1 show excellent agreement in both polarizations over a broad frequency range. In contrast, the $T_R$ or $T_L$ spectra from point 2 to point 6 in Figure~\ref{fig5}d-h show varying degrees of distortion. At turn-over frequencies, the dominant polarization (that due to the higher harmonic) of the folded spectra generally agree with the unfolded spectra down to ~25\% (~0.5~MK) of the peak brightness (~2~MK) for all except point 5 and 6, while the other polarization is good only up to about 50\% of the peak. This suggests that (not surprisingly) robust estimation of the coronal magnetic field strength may be limited to the brighter parts of the radio images, down to some limiting brightness temperature. We discuss this further in the next section. Because the spatial resolution continues to decrease below 3 GHz, spectra with sharp variations in frequency there tend to be distorted (e.g. the spectra of point 2 and 3 in RCP, and point 6 in both polarizations). For instruments with more antennas and longer baselines, such as the Jansky Very Large Array (VLA), the Chinese Solar Radioheliograph (CSRH), the upgraded Siberian Solar Radio Telescope (SSRT), or the future Frequency Agile Solar Radiotelescope (FASR), the observed spectra will more-closely approach the quality of the unfolded spectra of Figure~\ref{fig5}, and hence a spectrally-based analysis will be possible.  We discuss this further in section~\ref{level1}.

\subsection{Coronal Magnetography using an Image-Based (Level-0) Method}{\label{level0}
As described in section~\ref{comparison}, the gyroresonance spectrum is characterized by a sharp drop in brightness temperature from coronal to chromospheric values.  The spectral shape is merely a reflection of the fact that the temperature structure of the solar atmosphere along any LOS has a steep drop that occurs at the transition region, where the frequency-dependent gyroresonance layer drops out of the corona. Therefore, the magnetic field strength close to the base of the solar corona can be determined from the relevant harmonic of the gyrofrequency through equation (1), for lines-of-sight where gyroresonance emission dominates. Such a technique has been attempted using observational data to provide a rough estimate of magnetic fields at a few discrete levels corresponding to the fixed frequencies available for imaging in previous studies \citep{1998ApJ...501..853L}, and at a larger number of frequencies \citep{1994ApJ...420..903G,2011ApJ...728....1T,2010AstBu..65...60K}. However, as Figure~\ref{fig5} shows, for a realizable instrument with finite spatial resolution the observations compare best with the model-derived images only for the higher brightness temperatures (say above about 25\% of the peak brightness in the case of EOVSA).  This suggests use of an image-based approach. For simplicity, we refer to image-based analysis as the Level-0 method. In this section, we first introduce the principle of the Level-0 method, and apply it to the simulated EOVSA data. Then, we systematically evaluate its performance in deducing the magnetic field strength.

The Level-0 method uses three steps to deduce the coronal magnetogram, illustrated for three frequencies in Figure~\ref{fig7}: (i) for each radio map at a given frequency, the contour of a chosen brightness temperature is used to represent the ``edge'' of the gyroresonance surface of the presumed harmonic value $s=3$; (ii) the resulting magnetic field strength along this temperature contour can be deduced from the observing frequency used to produce the image, via equation (1), assuming $s=3$; (iii) a set of magnetic field contours, derived in the same way from multiple-frequency radio maps, can be overlaid to construct a final magnetogram at that chosen temperature level. Note that for each LOS, the brightness temperatures in RCP and LCP usually differ due to the dominant contributions being from different harmonics. To resolve this temperature ambiguity, we choose the dominant polarization, i.e. the one with the higher-frequency cut-off at the chosen threshold temperature, to correspond to the $X$ mode. Generally, the choice of harmonic $s=3$ is expected to be correct for the $X$ mode in coronal conditions in active regions, except over sunspots when the angle of the magnetic field to the LOS is close to $0^\circ$ or $180^\circ$ (this is usually only a significant factor for sunspots near disk center, as in our example).  In these regions, the lower optical depth causes the harmonic ratio to be 2:1.  Thus, we begin by making the usually correct assumption that $s=3$ for the $X$ mode, keeping in mind that in regions near the cores of sunspots away from the limb the correct choice may be $s=2$. Note that our assumption of $s=3$ as the highest optically-thick harmonic is the correct choice at points 1, and 4-6 in Figure~\ref{fig5}, but is not correct at points 2 and 3 where the harmonic ratio is 2:1 based on the unfolded spectra. Nevertheless, following the above steps, Figs.~\ref{fig7}a-c show the folded images in $X$ mode polarization at three frequencies. A relatively high threshold temperature of 0.72~$MK$ has been chosen to characterize the edge of the gyroresonance layer, as indicated by the dashed contour over each figure. Figure~\ref{fig7}d shows the corresponding magnetic field strengths along these overlaid contours. By applying the same technique for the many images of the data cube over the entire frequency range, we obtain the overall magnetogram shown in Figure~\ref{fig8}b.

We can evaluate the reliability of the Level-0 approach by comparing the Level-0-derived coronal field map, $B_{\rm fold}$ in Figure~\ref{fig8}b, with the actual magnetic field strength in the model, $B_{\rm model}$, at the height where the temperature $T_e=0.72~MK$ occurs, shown in Figure~\ref{fig8}a. The $B_{\rm fold}$ map extends over only the central part of the region, because the relatively high threshold temperature does not extend to the periphery. A dashed contour at 600~G, obtained from $B_{\rm model}$, is plotted over each figure for comparison. To provide a more quantitative comparison for each LOS, we calculate the percent error $(B_{\rm fold}-B_{\rm model})/B_{\rm model}$. Figure~\ref{fig8}c shows the resulting error distribution. Each color represents a certain error range, as indicated by the color bar. Above $\sim120$~G, both $B_{\rm model}$ and $B_{\rm fold}$ qualitatively agree very well over the active region where an error tolerance of  $<30$\% can be widely achieved. Note that abrupt variations from positive to large negative error, indicated by the blue and purple segments, occur in the regions near the sunspots, where the magnetic fields are nearly vertical (see point 2 in Figure~\ref{fig6}c), so that the harmonic ratio becomes 2:1 and our assumption of $s=3$ is violated. For quantitative comparison, Figure~\ref{fig8}d shows the $B_{\rm fold}$ (symbols) and $B_{\rm model}$ (solid line) sampled in a horizontal cut across the central part of the active region, as indicated by the dashed line in Figure~\ref{fig8}c. The profile of $B_{\rm fold}$, represented by asterisks, fits very well with the $B_{\rm model}$, except for the clearly discrepant, underestimated magnetic fields around the sunspots. A subjective correction of the discrepant points by a factor of 3/2, which comes from using a lower harmonic ($s=2$) in equation (1), shows improved agreement as indicated by open circles. In general, the result from the simple Level-0 procedure can fully exploit the $uv$-coverage of EOVSA to generate a reasonably reliable coronal magnetogram despite the limited spatial resolution of the folded data, although it is to be expected that there will be localized deviations near sunspots where the presumption of harmonic $s=3$ is inappropriate.

Although in principle we can derive a coronal magnetogram at any temperature level, we argued earlier based on Figure~\ref{fig5} that it is most favorable to choose a higher probing temperature to reduce the spectral distortion caused by two factors: the spatial resolution and the free-free emission, both of which can induce a gradual tail in the folded spectra above the turn-over frequency. It follows from equation (3) that the free-free emission depends on the square of the electron density $n_e$. As an illustration, we reran the model after increasing $n_e$ everywhere by a factor of 3 (i.e., an order of magnitude increase in free-free emission), with results shown in Figs.~\ref{fig8}e-f. In this case, the magnetic field $B_{\rm fold}$ derived from the Level 0 method deviates significantly below $\sim600$~G, corresponding to a turn-over frequency of 5 GHz at the third harmonic. This is due to much of the lower-field region being covered up by overlying free-free emission, exacerbated by the lower spatial resolution at lower frequencies. Finally, since the Level 0 magnetogram is derived at a constant temperature level, the downside of using a higher value for this threshold is that it corresponds to greater and more variable heights in the corona. Observations with better spatial resolution (e.g. VLA, CSRH or FASR) will support the choice of lower threshold temperatures (although in that case the spectrally-based ``Level-1'' approach may be better, as discussed in the next section).

Going back now to our original model with unenhanced density, Table 1 lists key quantities at our illustrative locations 1-6 in Figure~\ref{fig8}(c). Column (6) shows the percent error of magnetic field derived from the folded images, $B_{\rm fold}$, from Level 0 listed in column (4) with respect to the model, $B_{\rm model}$, in column (2). The total magnetic field at point 1 shows remarkable agreement because the turn-over frequency $\nu_{\rm fold}$ agrees well with $\nu_{\rm unfold}$ at the brightness temperature $0.72~MK$. At points 5 and 6, there are severe losses of spatial and spectral resolution in the low frequency range of the folded spectra, which results in a higher turn-over frequency and overestimated magnetic field strength $B_{\rm fold}$. The $B$ values at points 2 and 3 deviate from the model by $-34$\%, and $-16$\%, respectively, due to the use of an incorrect harmonic number. In real observations, an empirical correction to the choice of harmonic might be possible, but any automated approach will require experience with the actual data. In the next section, we further discuss the potential for more advanced, spectrally-based analysis, which we refer to as the Level-1 method, to correct for misdiagnosed harmonics.

\subsection{Coronal Magnetography using a Spectrum-Based (Level-1) Method}{\label{level1}}

We now briefly explore the prospects for obtaining more accurate coronal magnetograms through Level-1 analysis, which makes use of the spectra along each LOS in the two independent circular polarizations. Although EOVSA will not have the spatial resolution required to use this method in most cases, the VLA and some future instruments will. Along each LOS, the two distinct propagating electromagnetic modes, the $X$ mode and $O$ mode, can be utilized to remove ambiguity in the appropriate harmonic. As noted in section 4.1, it is commonly seen that the local spectra in $O$ mode fall off at a lower frequency than in the $X$ mode due to the dominance of different harmonics for the two modes. The Level 1 method can automatically detect a correct harmonic for each LOS in those cases where the steepest drops of the two polarized spectra occur, and use the corresponding turn-over frequencies $\nu_R$ and $\nu_L$ to match particular harmonic ratios (2:1 or 3:2, and possibly 4:3 in some cases). In general circumstances where $\nu_R\approx\nu_L$, a harmonic ratio of 3:3 is likely since the third harmonic then may be optically thick in both polarizations (e.g., when the viewing angle is nearly perpendicular to B). We test this approach to resolving the harmonic number in the absence of frequency-dependent resolution by applying our algorithm to the unfolded data cube. Figs.~\ref{fig9}a-b show the two examples of local spectra (obtained at LOS points 3 and 5) where the correct harmonic ratios, 3:2 and 4:3 respectively, have been successfully determined. Figure~\ref{fig9}c shows a percent error map, in which the Level 1 coronal magnetogram deduced from the unfolded image cube is compared with the $B_{model}$ map. Columns (3) and (7) in Table 1 represent the unfolded magnetic field derived from the Level-1 method, and the resulting percent error. Figure~\ref{fig9}d shows the corresponding harmonic map detected by the algorithm. It has been discussed in section 4.1 that $O$ mode might become totally optically thin at all observable frequencies (e.g. point 2), rendering the harmonic determination via the $O$ mode emission unusable. In Figure~\ref{fig9}d, this corresponds to the cores of the two sunspots, which are still presumed to have $s=3$ as the appropriate harmonic but are surrounded by the second-harmonic pixels. Therefore, it is reasonable to replace them with $s=2$. Point 2 has been manually corrected using $s=2$, yielding a percent error that reduces to $-3$\% (column (3) in Table 1). In general, the Level-1 method can resolve the misdiagnosed harmonics over the entire map, especially within the central active region. Figure~\ref{fig9}c shows that with higher resolution, systematic effects due to finite resolution can be significantly reduced to provide a more accurate result over a much larger area in the active region compared to Figure~\ref{fig8}c. Away from the active region, where $B$ is lower than a few hundred gauss, the unfolded magnetogram can still provide 30\% error on average. Since this algorithm relies heavily on the steepness of the polarized spectrum, it occasionally fails to diagnose higher harmonic ratios. For example, at point 5, the steepest slopes of both polarizations are detected at 2.3 GHz and 2.9 GHz (Figure~\ref{fig9}b), whose ratio is close to $s-1:s=4:5$. Instead, a harmonic ratio of 3:4 is clearly the correct one because the steep drop of $T_R$ at 1.5 GHz should be interpreted as the second harmonic. In brief, a more powerful algorithm could be developed to improve the Level 1 result as higher quality radio data become available in the future. We also anticipate that a forward fitting approach similar to that described for flaring loops in \citet{2013SoPh..288..549G} can be developed, which will have the advantage of providing additional coronal parameters from the gyroresonance spectra, such as density and angle of $B$ from the LOS. Such spectral fitting can also be used to explore whether alternative, non-Maxwellian particle distributions such as $\kappa-$ and $n-$ distributions exist in active regions \citep{2014ApJ...781...77F}.

\section{Discussion and Conclusions}

In this paper, we have demonstrated that EOVSA will be able to measure coronal magnetic field strengths in active regions, in areas where gyroresonance emission dominates. To evaluate the reliability of recovering the field strength, we performed the following steps: (1) We adopted an active region model, in which 3D coronal temperature, electron density and vector magnetic field are complex enough to provide an adequate challenge to test the feasibility of deriving magnetic field strength from microwave interferometric imaging. (2) We simulated a full set of unfolded images at multiple radio frequencies from 1-18 GHz, in which two dominating mechanisms, free-free emission and gyroresonance emission, have been included. (3) To provide a fair test of the resolving power of EOVSA, we embedded the active region into a realistic, frequency-dependent solar disk model. (4) A standard procedure was followed to generate two sets of model visibilities---those directly from the images (``unfolded'') and those folded through EOVSA's instrumental profile after 12 hours of rotation synthesis (``folded''). Sets of folded images were then reconstructed from the maximum entropy algorithm implemented in \emph{Miriad}.

We have used these reconstructed data cubes to investigate two alternative approaches to deriving coronal magnetic field maps, the simple Level-0 (image-based) method, and the more sophisticated Level-1 (spectrum-based) method.  We have shown that even with limited $uv$ coverage, in our case 78 baselines over a 12-hour observation, the Level-0 method can provide a reliable coronal magnetogram covering most areas within the active region. However, the deduced magnetic field strengths showed deviations from the model in a few areas near the cores of sunspots where the true harmonic in the $X$ mode was not the presumed $s=3$ value. To resolve the harmonic ambiguity, higher resolution radio data are required, although some ad hoc correction is possible simply by making the $s=2$ assumption in these sunspot areas (although this is influenced by the fact that we put the model active region near disk center so that the magnetic field over the umbrae is nearly parallel to the line of sight). We have briefly explored the power of the Level-1 method using direct model images of the unfolded data cube. The principle of this technique relies on a simple interpretation of the faithfully reconstructed polarized spectra, in which the $s:s-1$ harmonic frequency ratio can be determined in order to identify the relevant harmonic layer. The result has shown that most misdiagnosed harmonics have been corrected with significantly improved accuracy in a wider area.

The spectrum-based Level-1 method requires excellent spatial resolution as well as high image fidelity. With the advent of new radio facilities capable of multi-frequency solar radio imaging, such as EOVSA and the Jansky VLA, as well as future solar-dedicated instruments such as CSRH, the upgraded SSRT and FASR, the solar community is on the verge of having true coronal magnetograms that will grow increasingly reliable as analysis techniques are refined through experience. We expect that breakthroughs on long-standing solar topics, like the coronal structure of active regions, coronal heating and magnetic reconnection, can be achieved in the near future.

A final word about the interpretation of such coronal magnetograms is in order.  As for any remote sensing method for measuring magnetic fields, the magnetic field is measured at the location of emission formation rather than at a given height.  For photospheric lines, the range of heights is small and is generally ignored, but height variations become more important for chromospheric and coronal techniques such as the one described here.  As we have seen, the spatially-resolved gyroresonance radio spectrum provides the measurement of $T_b(\nu)$, which by virtue of the resonance ($\nu\rightarrow\nu_B\rightarrow B$) and optically-thick conditions ($T_b\rightarrow T$) becomes $T(B)$ (or equivalently $B(T)$).  By making use of the sharp drop in temperature that occurs at the base of the corona, we can constrain the height somewhat, but the precise height to which the measurement applies can vary by a few thousand km.  Therefore, coronal magnetograms will never be as straightforward to interpret as those for the photosphere, and their use will be most helpful in combination with 3D models.  This paper has shown that $B(T)$ can be reliably measured using radio interferometry data of solar active regions, which is the essential starting point for fruitful comparisons with 3D models.

\acknowledgments

This work was supported in part by NSF grants AST-1312802, AGS-1262772 and AGS-1250374, and NASA grants NNX11AB49G and NNX14AC87G to New Jersey Institute of Technology.



{\it Facilities:} \facility{EOVSA}.

\clearpage

\begin{table}[ht]
\caption{Comparison of derived parameters for the model, unfolded, and folded spectra, using Level 0 and Level 1. See detail description of each parameter in section 4.2-4.3}
\centering
\begin{tabular}{c c c c c c c c c}
\hline\hline
(1) & (2) & (3) & (4) & (5) & (6) & (7) \\
Point & $B_{model}$ (G) & $B_{unfold}$ (G) & $B_{fold}$ (G) & $s$ & $\Delta B^0_{\%}$ & $\Delta B^1_{\%}$ \\ [0.5ex]
\hline
1 & 1476 & 1422 & 1387 & 3 & -6.0  & -3.7  \\
2 & 1675 & 1667 & 1132 &(2)& -33.7 & -2.6  \\
3 & 997 & 1141 & 838  & 2 & -16.0 & 14.4  \\
4 & 394  & 411  & 385  & 3 & -2.1  & 4.3   \\
5 & 208  & 265  & 223  & 3 & 7.4  & 27.5  \\
6 & 134  & 152  & 160  & 3 & 19.1  & 12.8   \\ [1ex]
\hline

\tablecomments{The actual harmonics in column (5) are obtained from the Level 1 algorithm, except point 2 whose harmonic is manually corrected to minimize the percent error.}

\end{tabular}

\label{table:nonlin}
\end{table}

\begin{figure}
\plotone{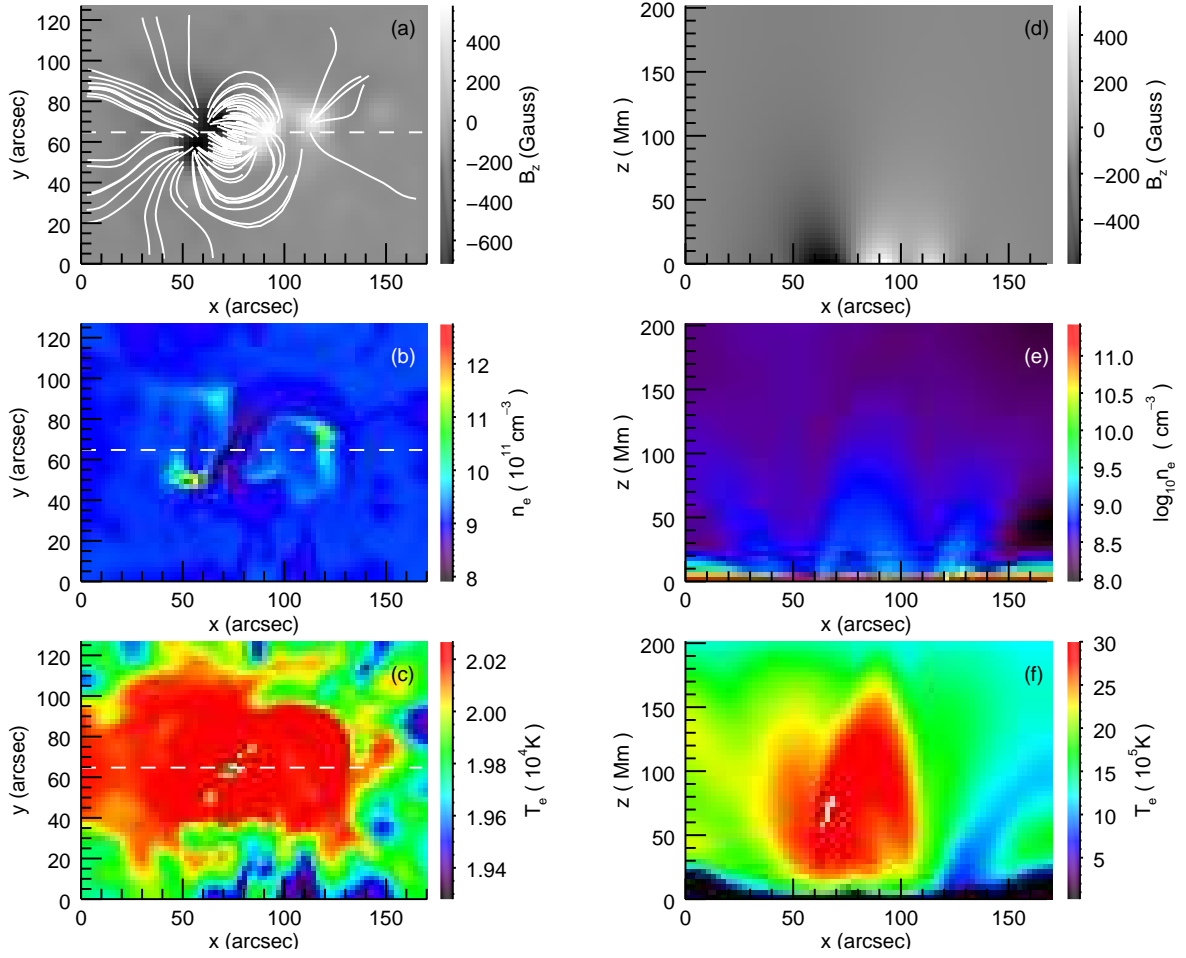}
\caption{(a)-(c) The longitudinal magnetic field $B_z$, electron density $N_e$ and temperature $T_e$ distribution at the model base corresponding to the transition region. The field of view is $172\arcsec\times127\arcsec$. (d)-(f) The corresponding height-dependent distribution of coronal parameters along the horizontal line in (a)-(c).\label{fig1}}
\end{figure}

\clearpage

\begin{figure}
\begin{center}
\includegraphics[width=7in]{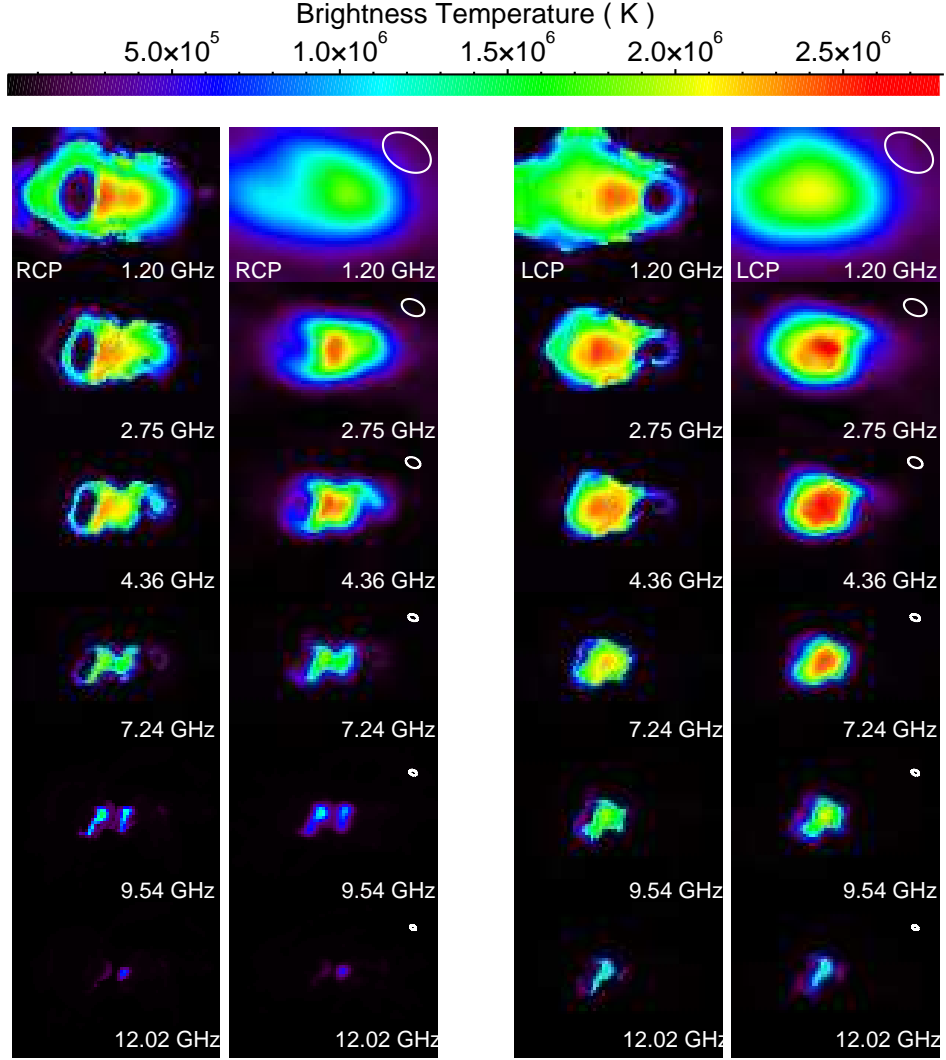}
\end{center}
\caption{ A comparison of the unfolded and folded images in both polarizations at six selected frequencies from 1-18 GHz. Images from left to right: the first and the third column are the unfolded images, which are derived from the radio emissions of the Mok's model, in RCP and LCP respectively; the second and the fourth columns are the corresponding restored images, after folding through EOVSA. The total FOV of each map is $172\arcsec$ in the $x$ direction, and $127\arcsec$ in the $y$ direction, with uniform spatial sampling of $2.4\arcsec$. \label{fig2}}
\end{figure}

\begin{figure}
\epsscale{.80}
\plotone{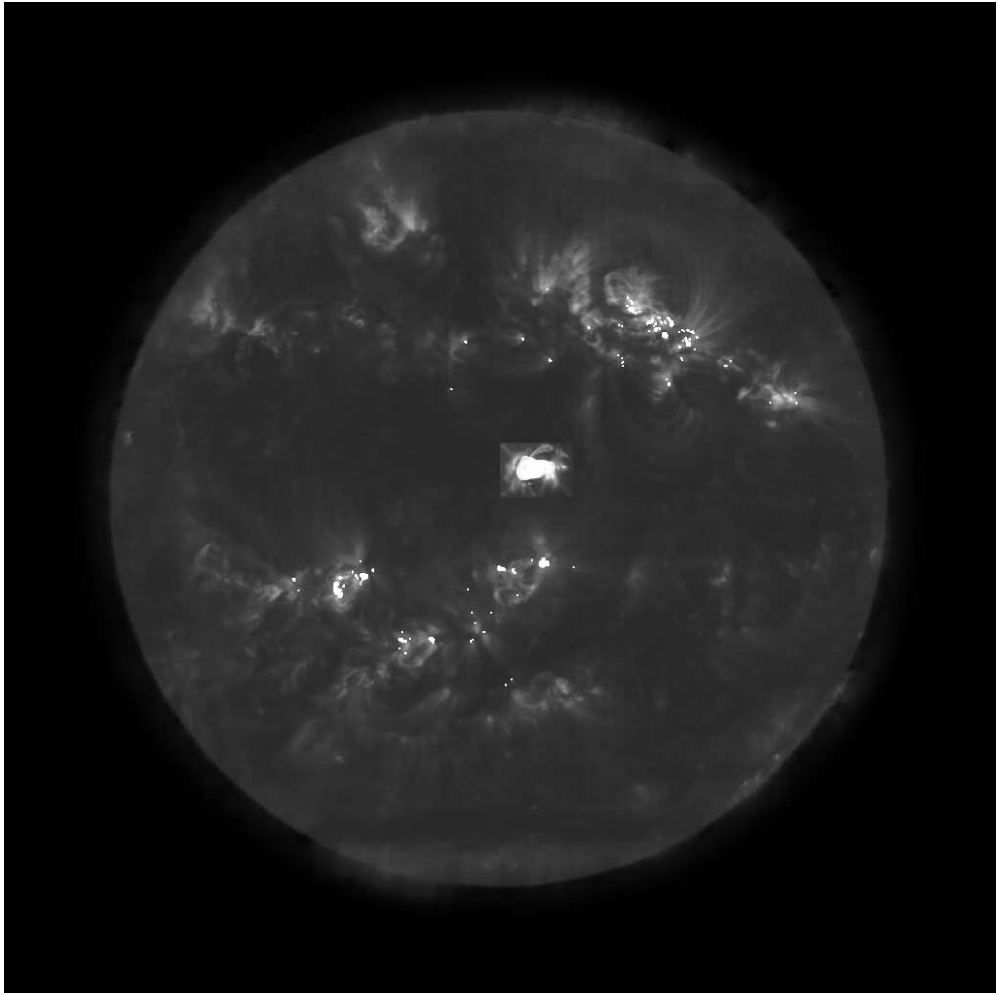}
\caption{A full-disk unfolded radio image in RCP, calculated for 4.15 GHz. The active region model is embedded near the center of the disk.\label{fig3}}
\end{figure}
\begin{figure}
\begin{center}
\includegraphics[width=3.0in]{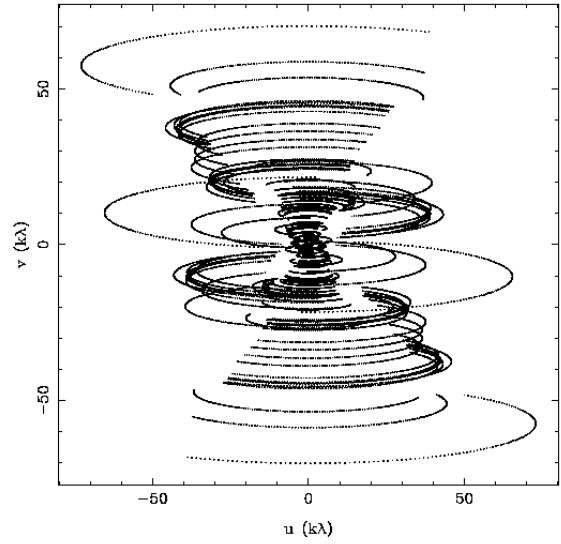}
\caption{The u,v coverage (EOVSA sampling function) at 18 GHz, which uses an hour angle range -6 hr \textless $h$ \textless 6 hr and +15$^\circ$ declination to produce the folded images.\label{fig4}}
\end{center}
\end{figure}

\begin{figure}
\begin{center}
\epsscale{1.3}
\plotone{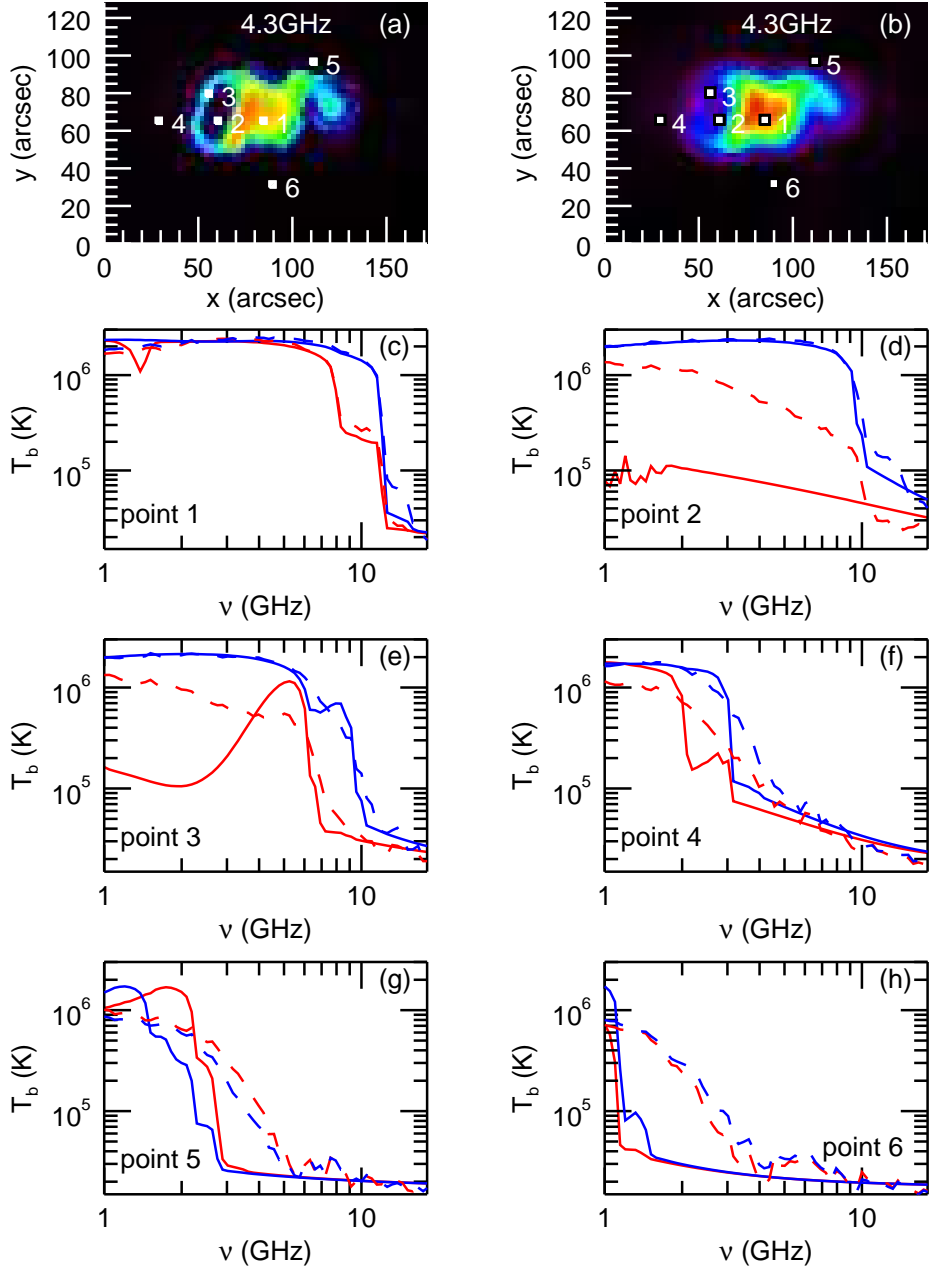}
\end{center}
\caption{(a)-(b) The RCP image at 4.34 GHz, before and after folding through the EOVSA instrument, respectively. (c)-(h) Comparison of the sampled spectra at six locations. The spectral curves in red and blue represent RCP and LCP, respectively. The solid lines, and dash lines with symbols represent the unfolded and folded spectra respectively. \label{fig5}}
\end{figure}

\begin{figure}
\begin{center}
\epsscale{1.0}
\plotone{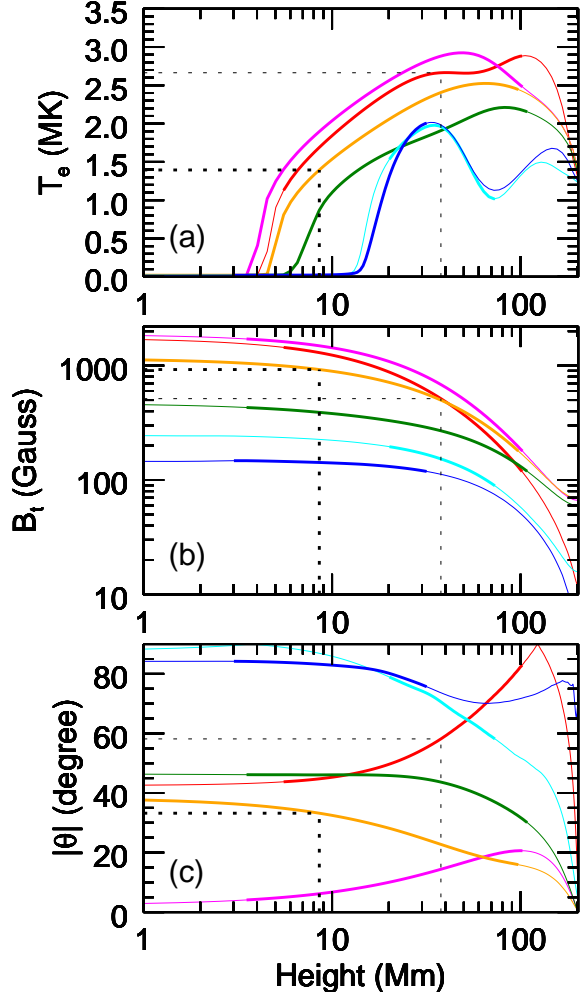}
\caption{Variations along the 6 lines of sight through the model of (a) the electron temperature $T_e$, (b) the magnitude of the total magnetic field $|B|$, and (c) its absolute angle $|\theta|$ as a function of coronal height.  The sampling points 1-6 are indicated in red, magenta, orange, green, cyan, and blue, respectively. The part of the height range corresponding to emission from 1 GHz to the turn-over frequency in the $T_X$ spectrum of each line of sight is accentuated by thicker lines. The pairs of dotted lines are discussed in the text.\label{fig6}}
\end{center}
\end{figure}

\begin{figure}
\begin{center}
\epsscale{1.0}
\plotone{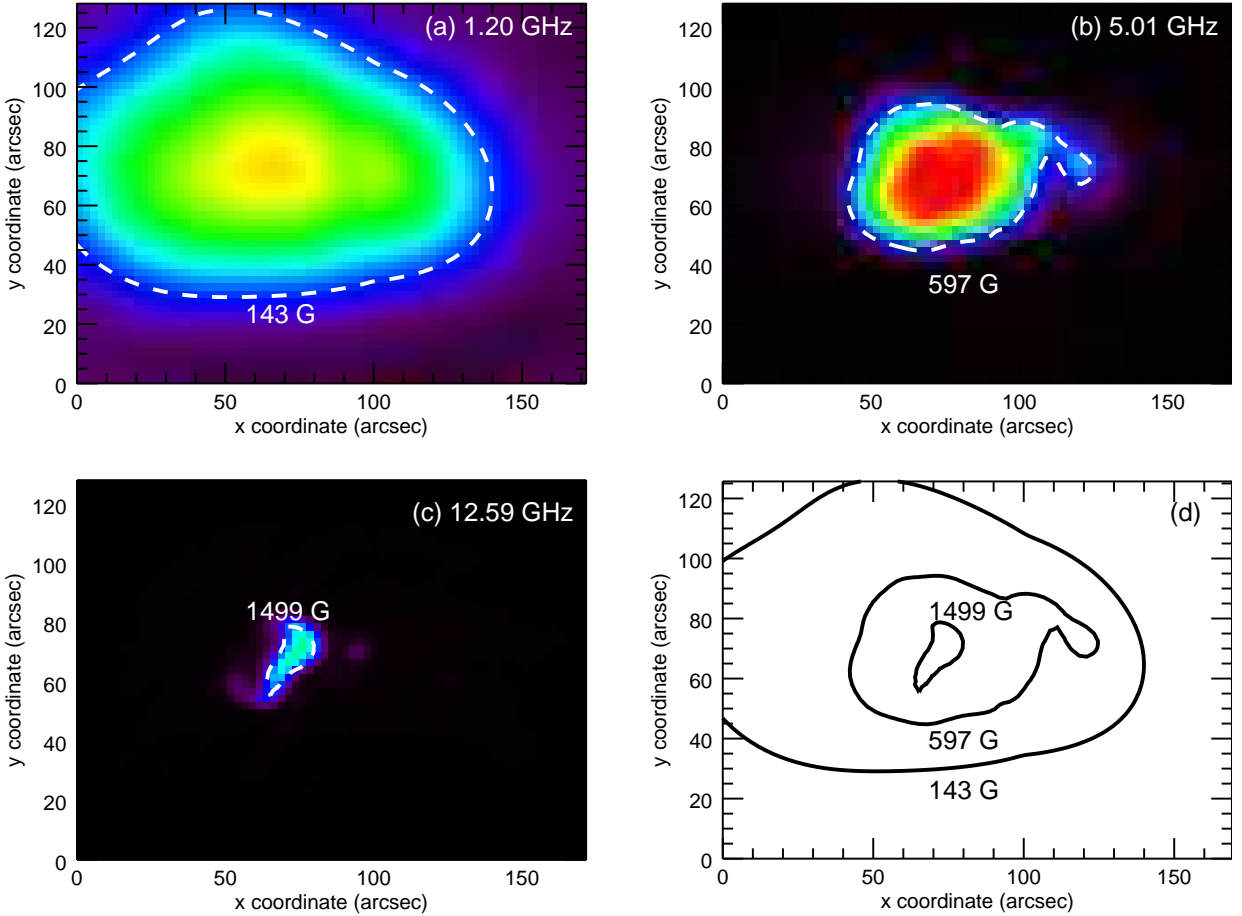}
\caption{(a)-(c) The brightness temperature maps at representative frequencies 1.20~GHz, 4.98~GHz and 12.47~GHz, respectively. The brightness temperature at each pixel shows the local value of either RCP or LCP brightness temperature, whichever is higher. A contour of a given brightness temperature $T_{b}\approx0.72~MK$, as represented by the dashed line, is plotted to outline the sharp edge of gyroresonance surface. The total magnetic fields along the contour are derived from equation (1). (d) The magnetic field contour map based on the three frequencies in 6(a)-(c) at $T_{b}\approx0.72~MK$. Similar contours at other frequencies fill out the final Level 0 coronal magnetic field map. \label{fig7}}
\end{center}
\end{figure}

\begin{figure}
\begin{center}
\includegraphics[width=2.48in]{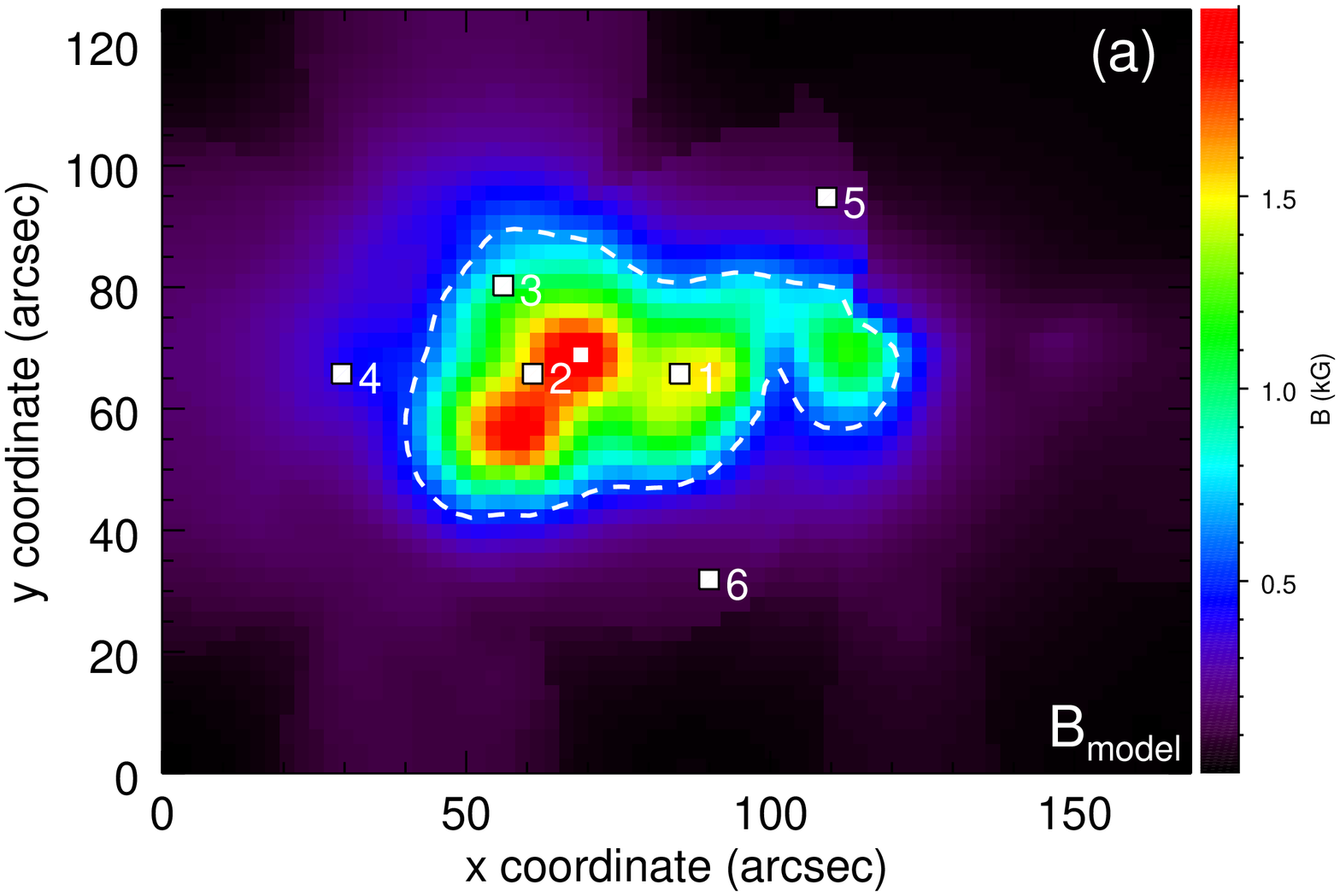}
\includegraphics[width=2.48in]{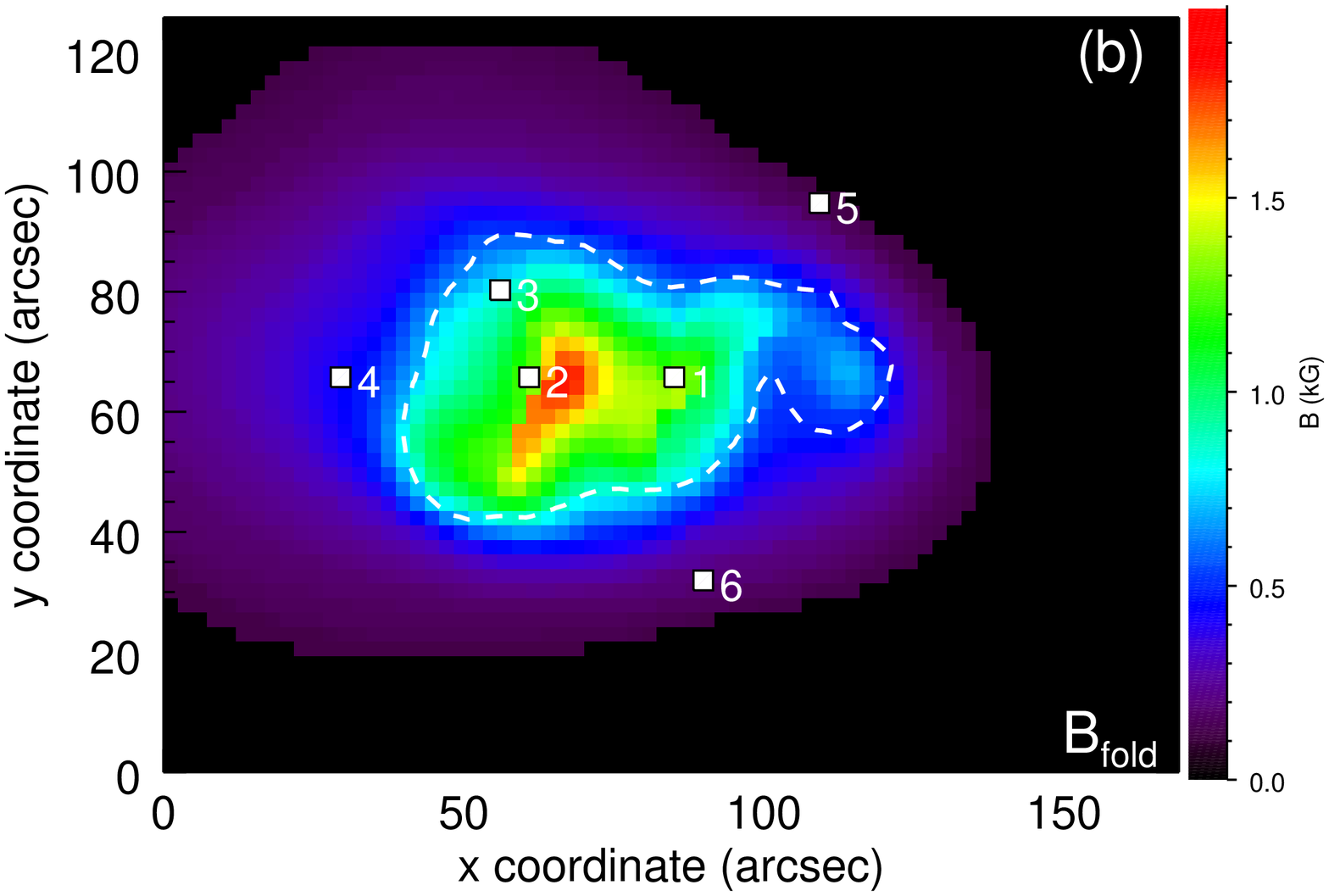}
\includegraphics[width=2.4in]{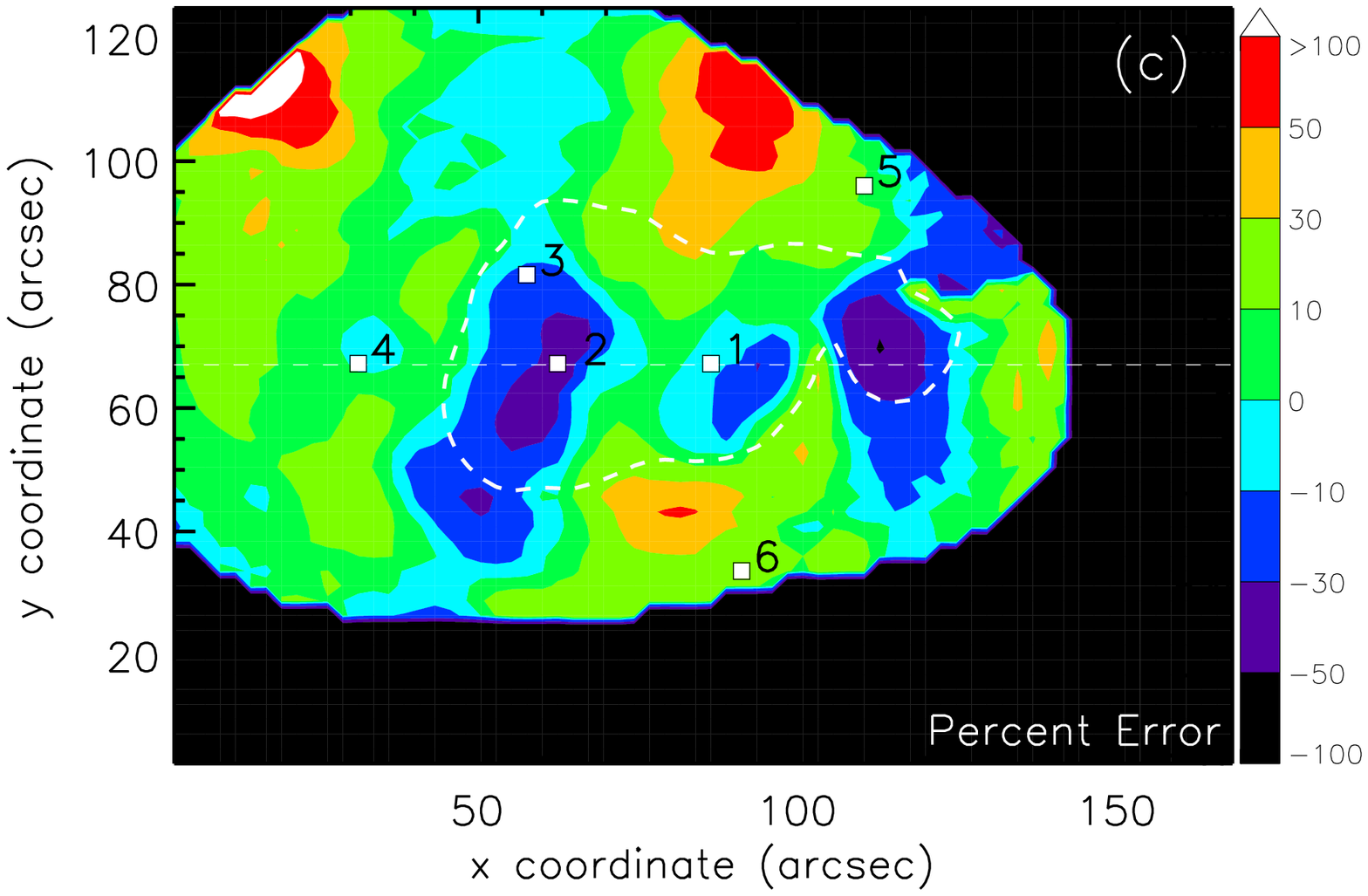}
\includegraphics[width=2.4in]{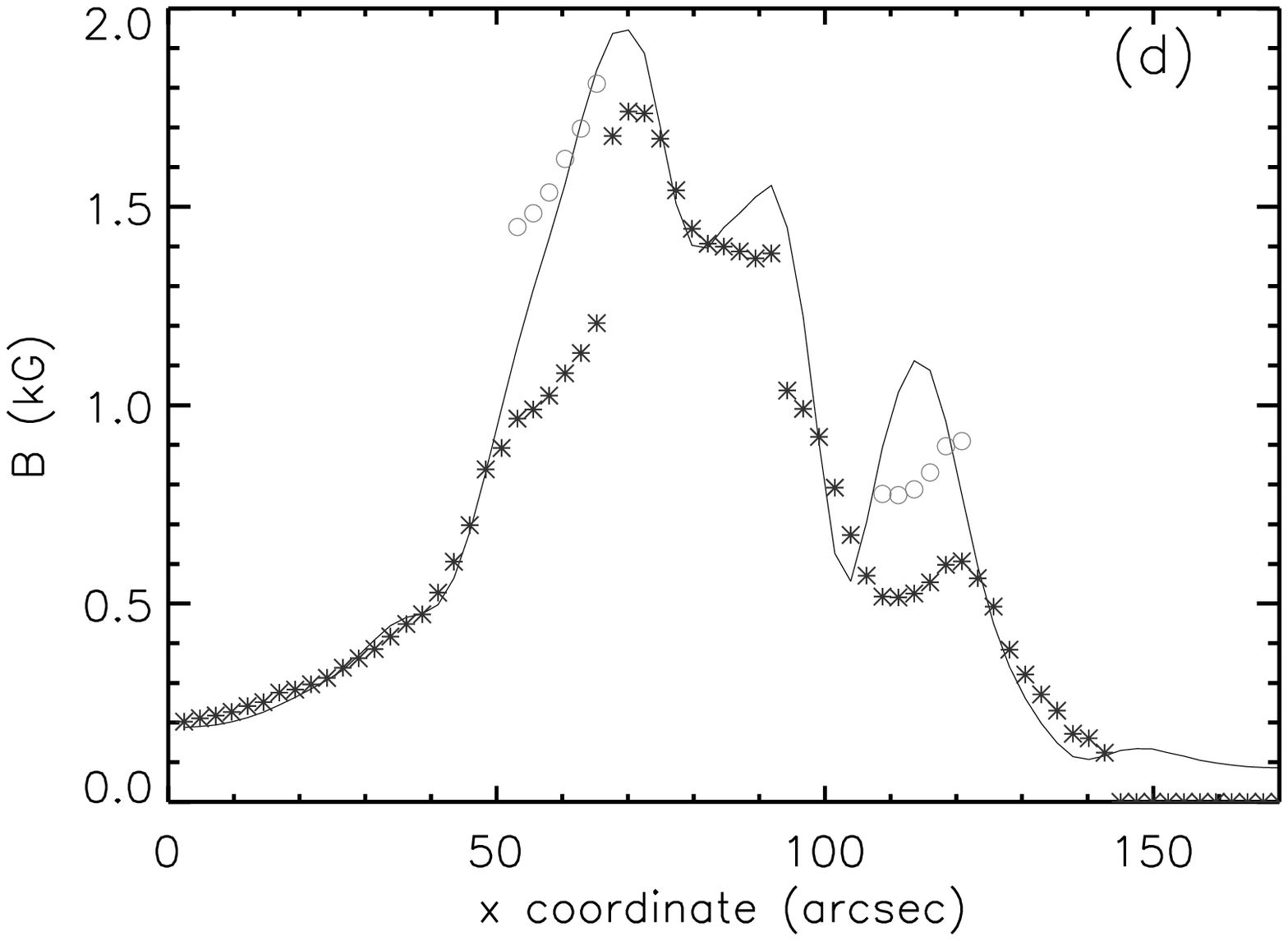}
\includegraphics[width=2.4in]{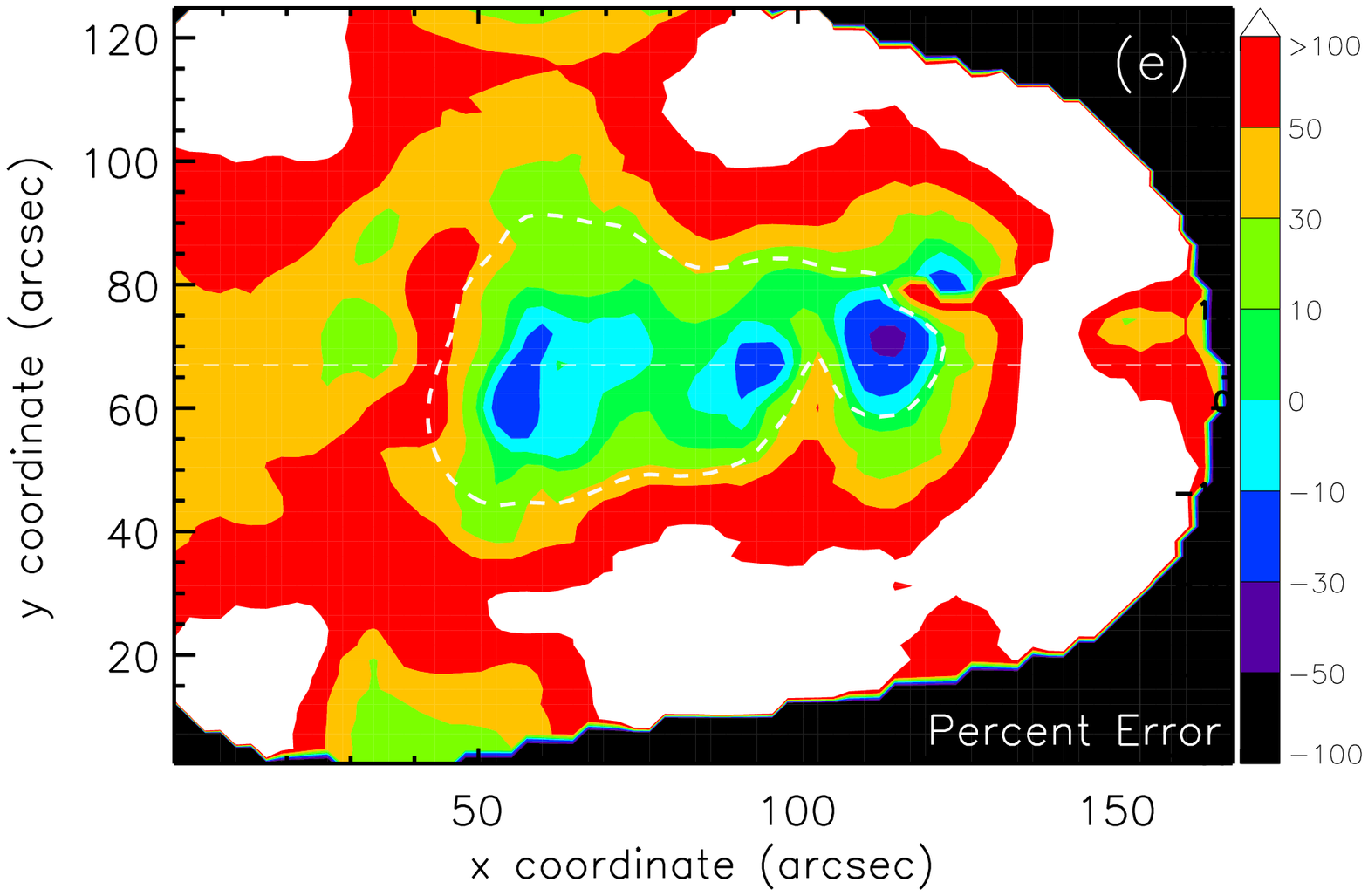}
\includegraphics[width=2.4in]{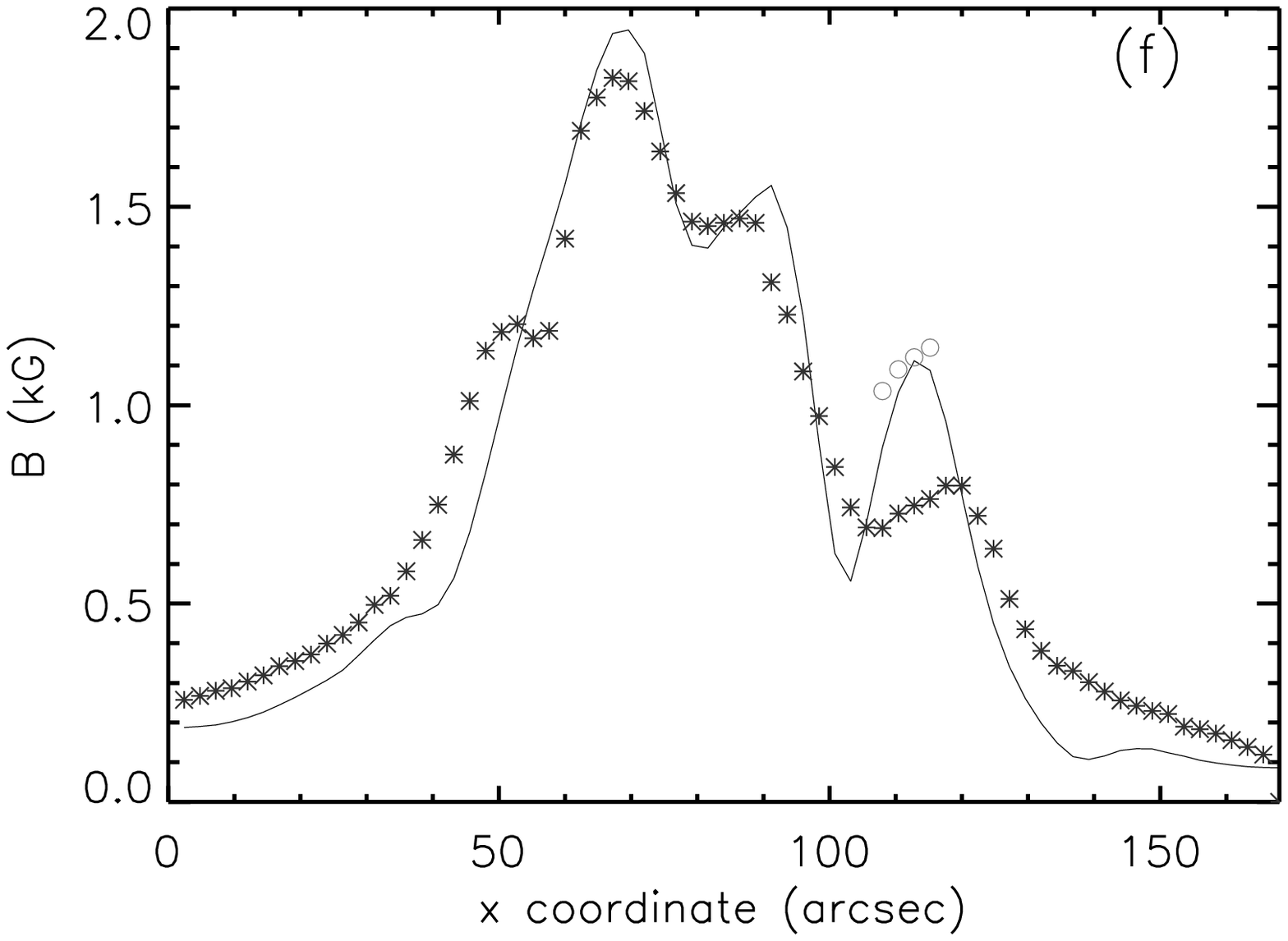}
\caption{(a) Result from the model: the magnetic field strength $B_{model}$ evaluated at coronal temperature $0.72~MK$. (b) Result from Level 0 method: the magnetic field map $B_{fold}$ at brightness temperature $0.72~MK$, which is derived from the 64-frequency folded data cube. (c) Schematic diagram of percent error in quantitative comparison of (b) with (a). (d) Comparison of $B_{model}$ (solid line) with $B_{fold}$ from the Level 0 method (asterisk symbols) along the horizontal line in Figure 8(c). The open circles represent manual corrections using a harmonic $s=2$. Figure 8(e) and (f) are obtained via similar procedures as Figure 8(c) and (d), but the electron density has been scaled to 3 time larger. The resulting strong free-free component reduces the accuracy widely for fields below 600 G.\label{fig8}}
\end{center}
\end{figure}

\begin{figure}
\begin{center}
\includegraphics[width=3.0in]{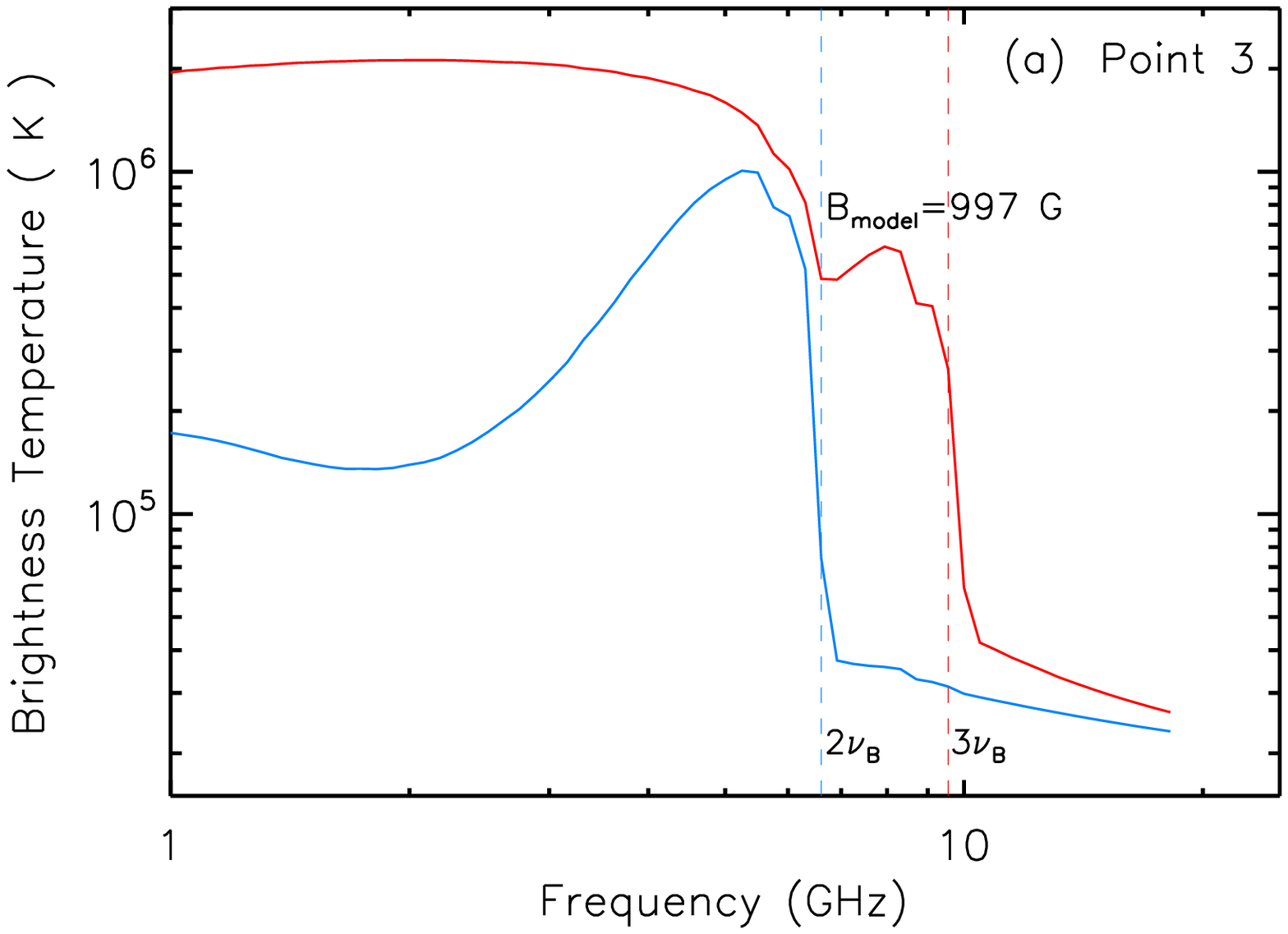}
\includegraphics[width=3.0in]{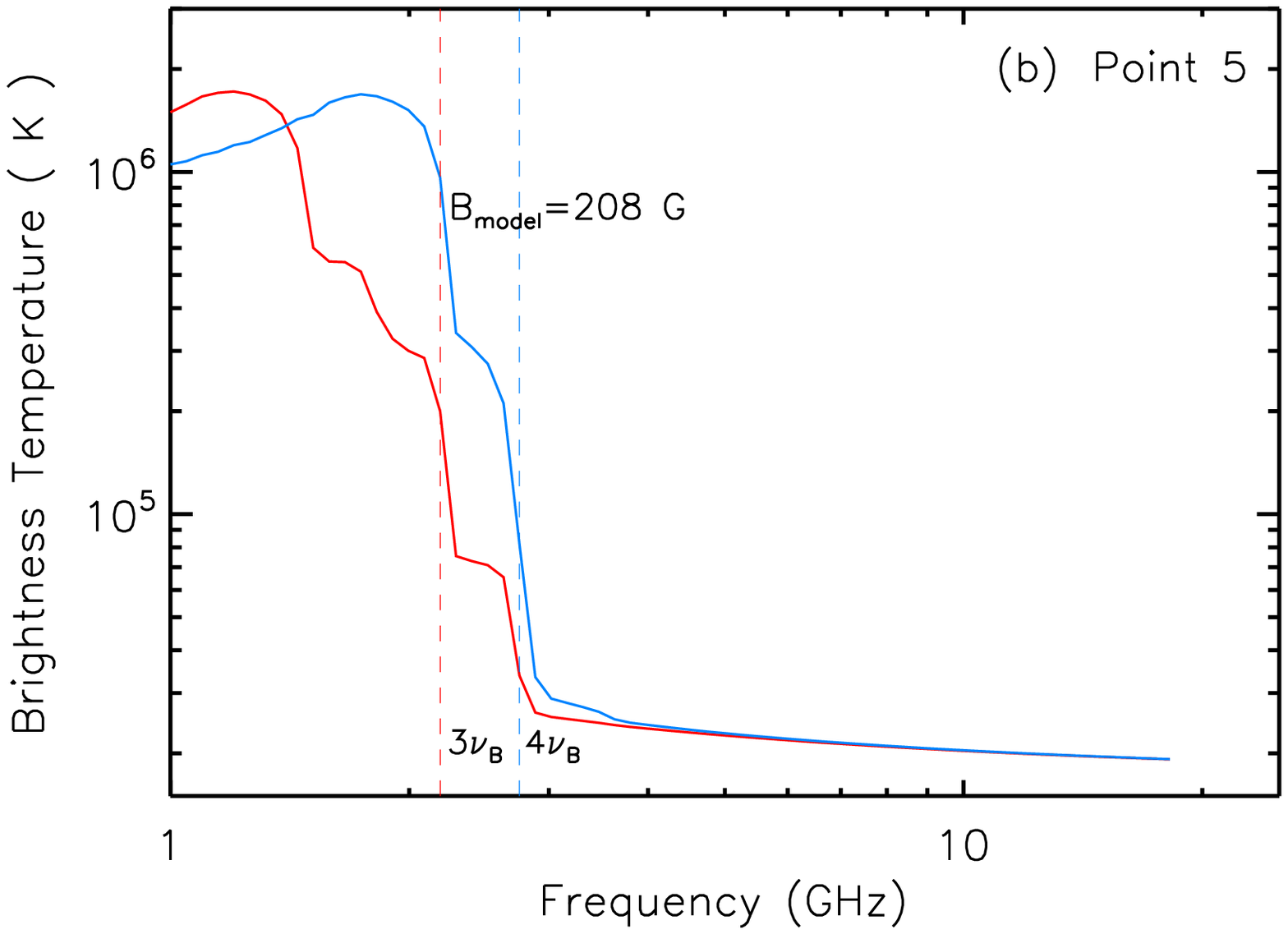}
\includegraphics[width=3.0in]{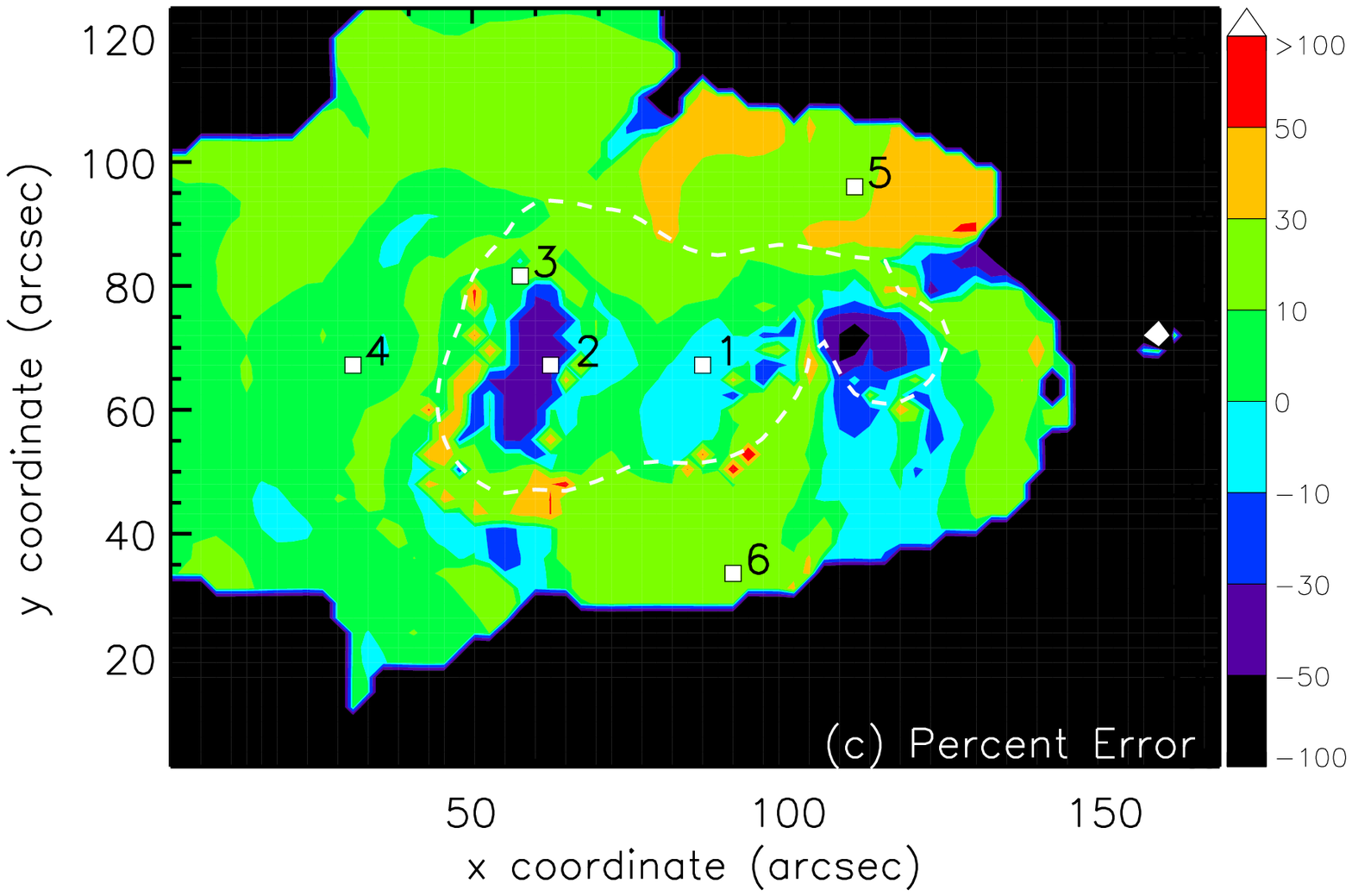}
\includegraphics[width=3.0in]{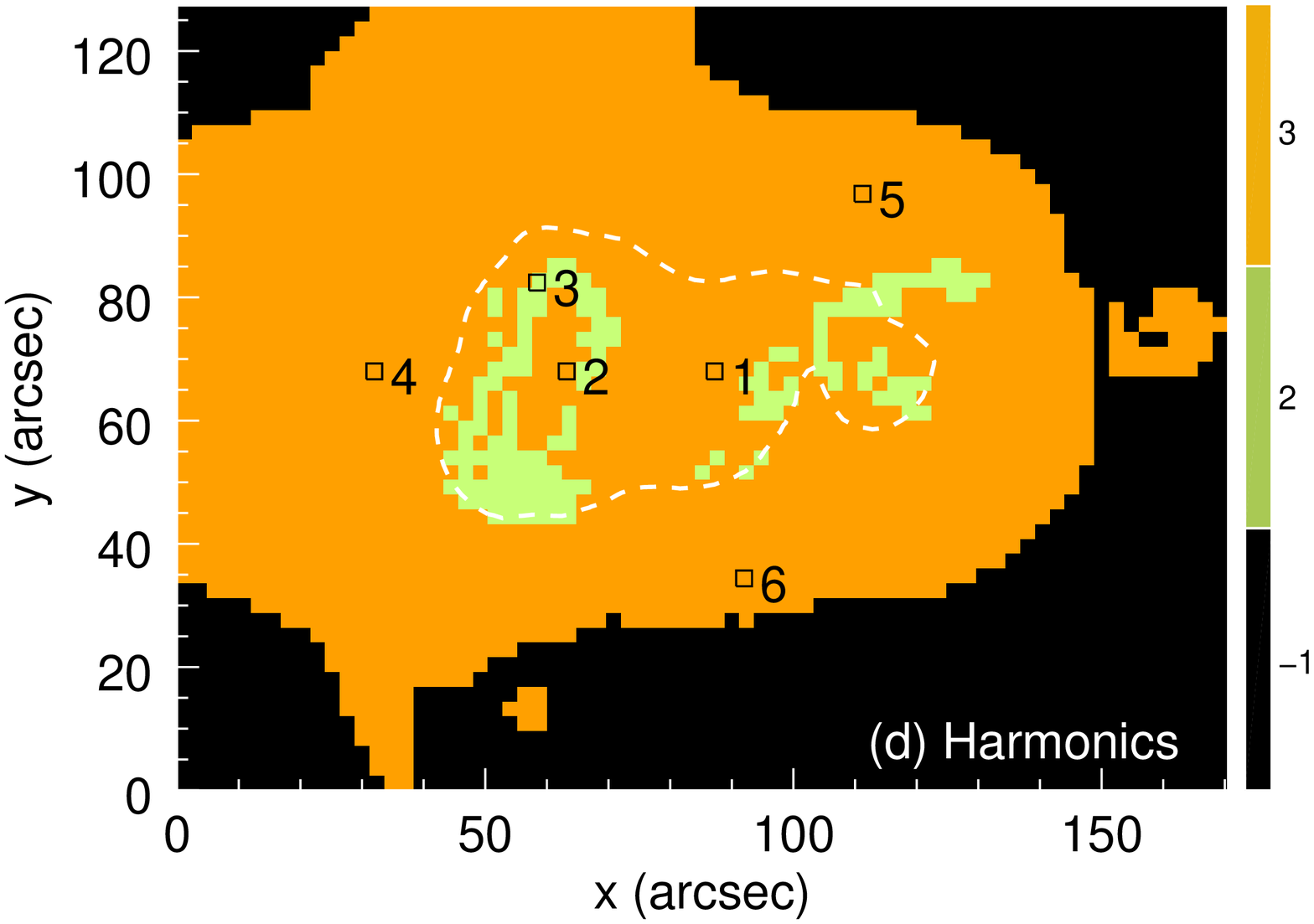}
\caption{(a)-(b) Two examples of polarized spectra from the unfolded data cube, where the gyrofrequencies, corresponding to the sharp drops in RCP (red) and LCP (blue), are found to match correct harmonic ratios. (c) The percent error map in which $B_{unfold}$, deduced from the Level 1 method, and $B_{model}$ are compared. Most misdiagnosed harmonics are automatically corrected within the central region. (d) The corresponding harmonic map diagnosed from the unfolded image datacube using the Level 1 algorithm. The black region, labeled $s=-1$ on the colorbar, shows the region where no harmonic could be undetermined, because $T_b$ is less than the probing temperature.\label{fig9}}
\end{center}
\end{figure}








\clearpage


{}




\end{document}